\documentclass[12pt,a4paper]{article}
\usepackage{jheppub}
\usepackage{amsmath,amsfonts,braket,slashed,amssymb,bm,bbm,psfrag,graphicx,color,dsfont,euscript}
\usepackage{mathtools}
\usepackage{multicol}
\usepackage{ctable}
\usepackage{subfigure}

\usepackage{hyperref}
\usepackage[toc,page]{appendix}

\allowdisplaybreaks
\newcommand{\df}{\mathrm{d}}

\newcommand{\X}{{\EuScript X}}

\newcommand{\G}{{\EuScript G}}

\newcommand{\nsl}{\rlap{\hspace{0.25mm}/}{n}}
\newcommand{\nbsl}{\rlap{\hspace{0.25mm}/}{\bar n}}
\newcommand{\spac}{{\hspace{0.3mm}}}
\newcommand{\braces}[1]{[\hspace{-0.5mm}[#1]\hspace{-0.5mm}]}
\newcommand{\dg}[0]{\Delta\Gamma_0}
\newcommand{\Dg}[0]{\Delta\Gamma_1}
\newcommand{\g}[0]{\gamma}
\newcommand{\pb}[0]{{\color{purple}\beta_0}}
\newcommand{\pbs}[0]{{\color{purple}\beta_0^2}}
\renewcommand{\r}[0]{\rho}
\renewcommand{\b}[0]{\beta_0}
\renewcommand{\a}[0]{\alpha_s}
\newcommand{\stoptocwriting}{\addtocontents{toc}{\protect\setcounter{tocdepth}{-5}}}
\newcommand{\resumetocwriting}{\addtocontents{toc}{\protect\setcounter{tocdepth}{\arabic{tocdepth}}}}

\begin{document}

\begin{titlepage}

\begin{flushright}
\normalsize
CERN-TH-2022-211\\
MITP/22-109\\ 
TUM-HEP-1441/22\\
December 20,2022
\end{flushright}

\vspace{1.0cm}
\begin{center}
\Large\bf\boldmath 
Factorization at Next-to-Leading Power and Endpoint Divergences in $gg\to h$ Production
\end{center}

\vspace{0.5cm}
\begin{center}
Ze Long Liu$^a$, Matthias Neubert$^{b,c}$, Marvin Schnubel$^b$ and Xing Wang$^d$\\
\vspace{0.7cm} 
{\sl ${}^a$Theoretical Physics Department, CERN, 1211 Geneva 23, Switzerland\\[3mm]
${}^b$PRISMA$^+$ Cluster of Excellence \& Mainz Institute for Theoretical Physics\\
Johannes Gutenberg University, 55099 Mainz, Germany\\[3mm]
${}^c$Department of Physics \& LEPP, Cornell University, Ithaca, NY 14853, U.S.A.\\[3mm]
${}^d$Excellence Cluster ORIGINS \& Physik Department T31\\
Technische Universität München, D–85748 Garching, Germany}
\end{center}

\vspace{0.8cm}
\begin{abstract}
 
We derive a factorization theorem for the Higgs-boson production amplitude in gluon-gluon fusion induced by a light-quark loop, working at next-to-leading power in soft-collinear effective theory. The factorization is structurally similar to that obtained for the $h\to\gamma\gamma$ decay amplitude induced by a light-quark loop, but additional complications arise because of external color charges. We show how the refactorization-based subtraction scheme developed in previous work leads to a factorization theorem free of endpoint divergences. We use renormalization-group techniques to predict the logarithmically enhanced terms in the three-loop $gg\to h$ form factor of order $\alpha_s^3\ln^k(-M_h^2/m_b^2)$ with $k=6,5,4,3$. We also resum the first three towers of leading logarithms, $\alpha_s^n\ln^{2n-k}(-M_h^2/m_b^2)$ with $k=0,1,2$, to all orders of perturbation theory.

\end{abstract}

\end{titlepage}

\tableofcontents
\newpage

\section{Introduction}
\label{sec:intro}

Factorization theorems are important for understanding observables sensitive to multiple energy scales. They provide a method for disentangling short-distance from long-range phenomena and allow for a resummation of large logarithmic corrections to all orders of perturbation theory. At leading order in scale ratios, a typical factorization theorem consists of a product or a convolution of functions that are each associated with a single scale. At subleading power, however, several complications arise. With the upcoming analysis of the Run-3 dataset of the large hadron collider (LHC) at CERN, it will be possible to measure the properties of the Higgs boson with unprecedented precision. It is, therefore, necessary to have equally precise theoretical predictions at hand. The main production channel for the Higgs boson is the gluon-gluon fusion process, $gg\to h$, mediated via quark loops. The top quark gives the largest contribution, and its effects have been studied up to three-loop order \cite{Czakon:2020vql}. While this contribution is purely short-distance dominated, the subleading contributions from light quarks are sensitive to three very different mass scales, $M_h\gg\sqrt{M_h m_b}\gg m_b$, where here and below we focus on the case of a $b$-quark loop. Estimates for the impact of this contribution vary in the range between 9\,--\,13\%, depending on whether one takes the value for the $b$-quark pole mass $m_b^{\rm pole}\approx 4.8$\,GeV \cite{ParticleDataGroup:2022pth} or the running mass $m_b(M_h)\approx 2.6$\,GeV \cite{Aparisi:2021tym}. In order to reduce this ambiguity, it is crucial to resum large logarithmic contributions in the scale ratio $M_h/m_b$ to all orders of perturbation theory. The leading such terms are of order $\alpha_s^n\ln^{2n}(-M_h^2/m_b^2)$. The goal of this work is to derive a factorization theorem for this process, based on which this resummation can be accomplished.

In \cite{Liu:2019oav,Liu:2020wbn,Liu:2020tzd}, we have applied advanced methods of soft-collinear effective theory (SCET) \cite{Bauer:2000yr,Bauer:2001yt,Bauer:2002nz,Beneke:2002ph,Becher:2014oda} to derive the corresponding factorization theorem for the Higgs-boson decay $h\to\gamma\gamma$ mediated by a $b$-quark loop. This was the first complete SCET factorization formula for an observable that is of next-to-leading power (NLP) in small scale ratios. Compared with the contribution of the top quark, the Higgs coupling to bottom quarks provides the power suppression in the expansion parameter $\lambda\sim m_b/M_h$. It is by now well-known that scale factorization at NLP is full of complexities. The factorization theorems contain a sum over convolutions of Wilson coefficients with operator matrix elements, which are plagued by endpoint singularities. They manifest themselves in divergent convolution integrals over products of component functions \cite{Ebert:2018gsn,Beneke:2019kgv,Moult:2019mog,Moult:2019uhz,Beneke:2019oqx,Moult:2019vou,Liu:2019oav,Wang:2019mym,Beneke:2020ibj,Liu:2020wbn,Liu:2020tzd,Beneke:2022obx}. One may interpret such divergences as a failure of dimensional regularization and the $\overline{\text{MS}}$ subtraction scheme, because some poles in the dimensional regulator are not removed by renormalizing the individual component functions, and hence naive scale separation is violated. Standard tools are then insufficient to obtain well-defined, renormalized factorization theorems. 

This work is dedicated to generalizing the methodology developed for the factorization of the light-quark induced contribution to the $h\to\gamma\gamma$ decay amplitude (to which we will from now on refer to as the ``photon case'') to the non-abelian counterpart, the fusion process $gg\to h$ via light-quark loops (below often referred to as the ``gluon case''). Following closely the steps laid out in our previous works, we will derive the bare factorization theorem in section~\ref{sec:derivfacto}, pointing out important differences with respect to the photon case, which result from the fact that the external gluons carry color. We show how implementing the refactorization-based subtraction (RBS) scheme developed in \cite{Liu:2019oav,Liu:2020wbn} makes it possible to write down a factorization theorem that is free of endpoint divergences. In section~\ref{sec:reno}, we discuss the renormalization of the factorization theorem. While the renormalization of the component functions and the regularization of endpoint divergences in the RBS scheme do not commute, we show that it is possible to absorb all additional ``mismatch contributions'' into a redefinition of one of the hard matching coefficients in the factorization formula. Sections~\ref{sec:RG} and \ref{sec:LL} are dedicated to deriving the renormalization-group (RG) evolution equations for all entities in the factorization theorem, and using them to predict the leading large logarithmic terms in in the three-loop $gg\to h$ amplitude, respectively. In section~\ref{sec:resum}, we solve the evolution equations in RG-improved perturbation theory and resum the infinite towers of logarithms $\alpha_s^n\ln^{2n-k}(-M_h^2/m_b^2)$ with $k=0,1,2$ to all orders of perturbation theory. We conclude in section~\ref{sec:con}. Several technical details of our calculations are collected in four appendices.

\section{Derivation of the factorization theorem}
\label{sec:derivfacto}

In this section, we apply SCET to disentangle the relevant energy scales and obtain a factorization formula for the light-quark induced $gg\to h$ production amplitude, following closely our previous work on the corresponding contributions to the $h\to\gamma\gamma$ decay amplitude \cite{Liu:2019oav,Liu:2020wbn}. In the following, we first introduce some basic notions and SCET and illustrate the main challenges faced when applying SCET at NLP in scale ratios. We then point out the main differences in the treatment of the $gg\to h$ and $h\to\gamma\gamma$ amplitudes, which arise due to the fact that the external gluons carry color and hence are unphysical external states.

\subsection{General remarks about SCET at next-to-leading power}

Much of the power of SCET derives from the fact that it allows one to factorize hard, collinear, and soft interactions already at the Lagrangian level (at leading power). Extending the formalism to NLP, however, reintroduces interactions between the different sectors. It is a highly non-trivial task to ensure that scale separation still works in higher orders in power counting. We use $\lambda=m_b/M_h$ as the expansion parameter of SCET. As usual in SCET, we decompose all momenta into light-cone components
\begin{equation}
	\label{eq:momdecomp}
	\ell^\mu=(n_1\cdot\ell)\frac{n_2^\mu}{2}+(n_2\cdot\ell)\frac{n_1^\mu}{2}+\ell_\perp^\mu\,.
\end{equation}
Here, $n_1$ and $n_2$ are two light-like reference vectors aligned with the directions of the external gluons, i.e.\ $n_i\parallel k_i$, which satisfy $n_i^2=0$ and $n_1\cdot n_2=2$. In the rest frame of the Higgs boson, they can be chosen as $n_1^\mu=(1,0,0,1)$ and $n_2^\mu=(1,0,0,-1)$. In the following, we will often use the conjugate vectors $\bar{n}_1^\mu\equiv n_2^\mu$ and $\bar{n}_2^\mu\equiv n_1^\mu$. Indicating the scalings of the individual momentum components as $(n_1\cdot\ell,n_2\cdot\ell,\ell_\perp)$, we find that the following modes are relevant in the low-energy effective theory:
\begin{equation}
	\label{eq:mom-regions}
	\begin{aligned}
		\text{hard}(h):&\qquad\ell^\mu\sim(1,1,1)M_h \,, \\
		n_1\text{-collinear}(c):&\qquad\ell^\mu\sim(\lambda^2,1,\lambda)M_h \,, \\
		n_2\text{-collinear}(\bar{c}):&\qquad\ell^\mu\sim(1,\lambda^2,\lambda)M_h \,, \\
		\text{soft}(s):&\qquad\ell^\mu\sim(\lambda,\lambda,\lambda)M_h \,.
	\end{aligned}
\end{equation}
Matching the Standard Model (SM) onto the effective theory is a two-step process, $\text{SM}\to\text{SCET}_1\to\text{SCET}_2$. In the intermediate effective theory SCET$_1$ exchanges between the soft and collinear sectors are still present, and one needs hard-collinear modes obeying the scaling relations
\begin{equation}
	\begin{aligned}
		n_1\text{-hard-collinear }(hc):&\qquad\ell^\mu\sim(\lambda,1,\lambda^\frac12)M_h \,, \\
		n_2\text{-hard-collinear }(\overline{hc}):&\qquad\ell^\mu\sim(1,\lambda,\lambda^\frac12) M_h \,.
	\end{aligned}
\end{equation}
Integrating out the hard-collinear modes results in the so-called radiative jet functions as matching coefficients \cite{Moult:2019mog,Beneke:2019oqx,Liu:2020ydl,Liu:2021mac}. In SCET, operators are built of so-called gauge-invariant (hard-)collinear building blocks, which are composite objects invariant under collinear gauge transformations. This provides the advantage that gauge invariance is explicit despite the fact that SCET is intrinsically non-local through the appearance of Wilson lines. 

A common feature of NLP SCET problems is the occurrence of endpoint-divergent convolution integrals. Some of them can be regularized using dimensional regularization, while others require additional analytic (or rapidity) regulators \cite{Becher:2010tm,Becher:2011dz,Chiu:2011qc,Chiu:2012ir}. Even though the dependence on the analytic regulator cancels in the sum of all terms in the factorization formula, the presence of endpoint singularities upsets the usual renormalization in the $\overline{\text{MS}}$ scheme, because renormalizing the composite operators and Wilson coefficients in the effective theory does not remove all divergences. This is the bottleneck of all NLP problems. The refactorization-based subtraction (RBS) scheme proposed in \cite{Liu:2019oav,Liu:2020wbn} addresses this problem in a systematic way. Based on exact $d$-dimensional refactorization conditions, it exploits the fact that the integrands of the divergent integrals in different terms in the factorization theorem become identical in the singular regions. This allows for a rearrangement, which removes the endpoint divergences. The importance of refactorization conditions and refactorization-based subtractions has also been emphasized in later work \cite{Beneke:2020ibj,Beneke:2022zkz}, and it is the only known systematic method to deal with factorization at NLP.

\subsection[Factorization in $h\to\gamma\gamma$ decay]{\boldmath Factorization in $h\to\gamma\gamma$ decay}
\label{subsec:hgaga}

\begin{figure}
	\centering
	\includegraphics[width=\textwidth]{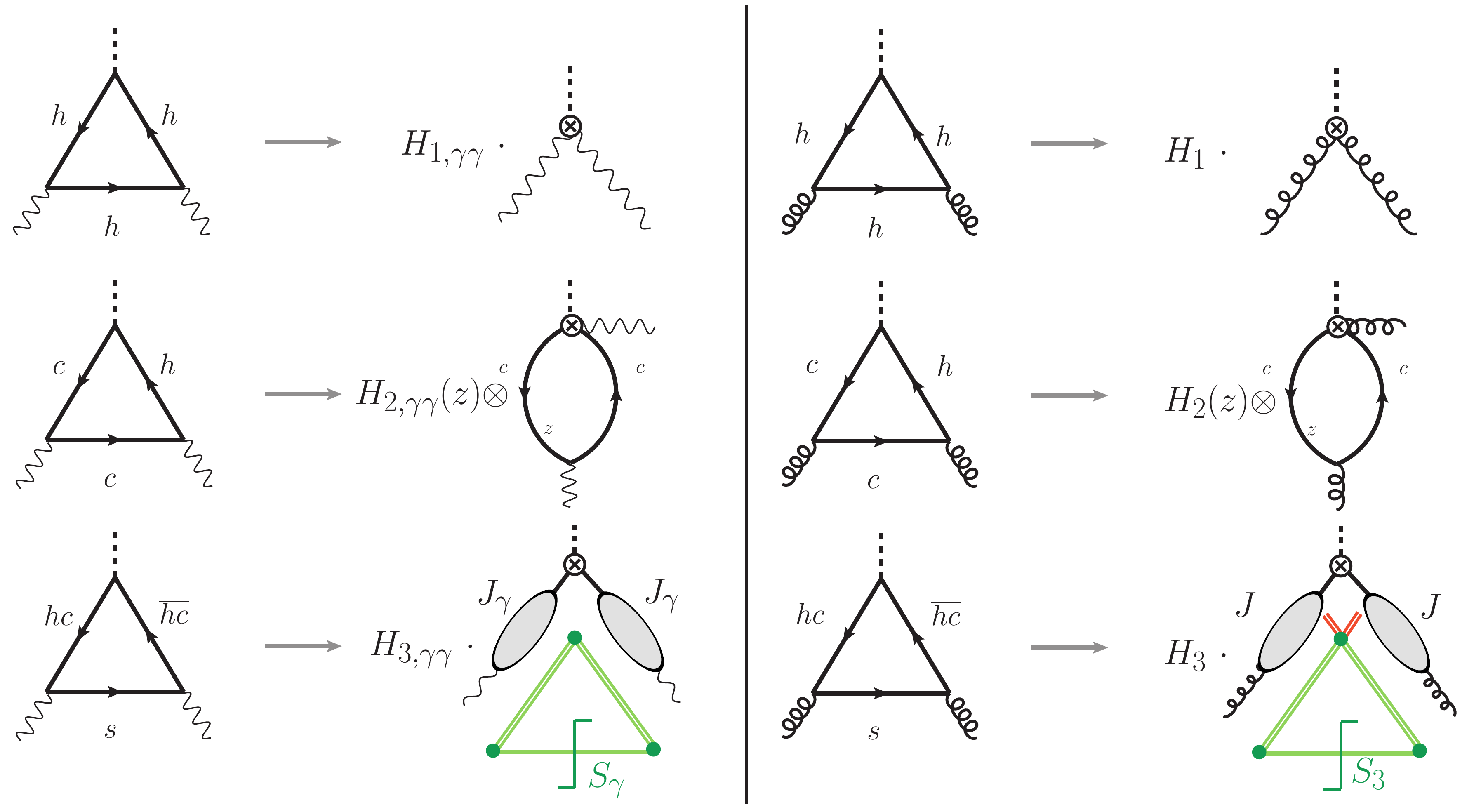}
	\caption{Relevant regions of loop momenta contributing to the amplitudes for $h\to\gamma\gamma$ (left) and $gg\to h$ (right). The convolution symbol $\otimes$ in the second term means an integral over the momentum-fraction variable $z$. The green double lines in the third term represent finite Wilson-line segments, whereas the red double lines indicate semi-finite Wilson lines in the adjoint representation of $SU(N_c)$, which are present only for the gluon case.}\label{fig:fact}
\end{figure}

Before studying the factorization properties of the $gg\to h$ production process, we find it instructive to recapitulate the main steps in the derivation of the analogous factorization theorem for the $h\to\gamma\gamma$ decay amplitude. We begin with the factorization formula in terms of bare Wilson coefficients and operator matrix elements derived in \cite{Liu:2019oav}. It consists of the matrix elements of three bare SCET operators $O_{i,\gamma}^{(0)}$ multiplied (or convoluted) with bare Wilson coefficients $H_{i,\gamma}^{(0)}$, which account for the hard matching corrections arising when the full theory (i.e., the SM with the top quark integrated out) is matched onto SCET. The factorization theorem reads
\begin{equation}
	\label{eq:factorizationPhoton}
	\mathcal{M}_{b}(h\to\gamma\gamma)=H_{1,\gamma}^{(0)}\langle O_{1,\gamma}^{(0)}\rangle+2\int_0^1\df z\,H_{2,\gamma}^{(0)}(z)\langle O_{2,\gamma}^{(0)}(z)\rangle+H_{3,\gamma}^{(0)}\langle O_{3,\gamma}^{(0)}\rangle \,.
\end{equation}
The three terms correspond to different regions of loop momenta contributing to the decay amplitude. The situation is portrayed in figure~\ref{fig:fact} for both the $h\to\gamma\gamma$ (left) and $gg\to h$ (right) process. A region analysis of the full-theory one-loop Feynman diagram reveals that the momentum flowing through the propagator connecting the two gauge bosons can be either hard, $n_i$-collinear or soft. The same regions are also relevant for multi-loop graphs. The first term in the factorization theorem is obtained when all loop momenta are hard. In the effective theory, the loop is then shrunken to a point-like interaction connecting a Higgs field to two gauge fields, describing photons moving along the light-like directions $n_1$ and $n_2$. The second term arises when the loop momentum is collinear with one of the photon directions. The operator $O_{2,\gamma}^{(0)}(z)$ contains a Higgs field, an $n_2$-collinear photon field, and two $n_1$-collinear $b$-quark fields, which annihilate each other to produce the photon moving along the direction $n_1$. The variable $z\in[0,1]$ indicates the fraction of the photon momentum carried by the $n_1$-collinear quark. Interchanging the photon directions $n_1$ and $n_2$ yields the same result, hence giving rise to the factor~2 in the factorization formula. The third term arises when the loop momentum is soft, which forces the other two quark propagators to be hard-collinear. Formally, the operator $O_{3,\gamma}^{(0)}$ contains the time-ordered product of the scalar Higgs current with two insertions of the subleading-power SCET Lagrangian \cite{Beneke:2002ph}, in which hard-collinear fields are coupled to a soft quark field. Integrating out the hard-collinear fields, the matrix element of this third operator can be further factorized into the double-convolution of two radiative jet functions and a soft-quark soft function, i.e.\ \cite{Liu:2019oav}
\begin{equation}
	\label{eq:O3factoPhoton}
	\begin{aligned}
       \langle O_{3,\gamma}^{(0)}\rangle 
       &= \frac{\varepsilon_1^\perp(k_1)\cdot\varepsilon_2^\perp(k_2)}{2}\int_0^\infty\frac{\df \ell_+}{\ell_+} 
        \int_0^\infty\frac{\df \ell_-}{\ell_-} \\
       &\quad\times \left[J_\gamma^{(0)}(M_h\ell_+)J_\gamma^{(0)}(-M_h\ell_-)
        +J_\gamma^{(0)}(-M_h\ell_+)J_\gamma^{(0)}(M_h\ell_-)\right]  S_\gamma^{(0)}(\ell_+\ell_-) \,,
	\end{aligned}
\end{equation}
where $\varepsilon_i^\perp(k_0)$ denote the (transverse) photon polarization vectors, while $J_\gamma$ and $S_\gamma$ are the radiative jet and soft functions, respectively. The properties of these functions have been studied in great detail in \cite{Liu:2020ydl,Liu:2020eqe}.

\begin{figure}
	\centering
	\includegraphics[width=\textwidth]{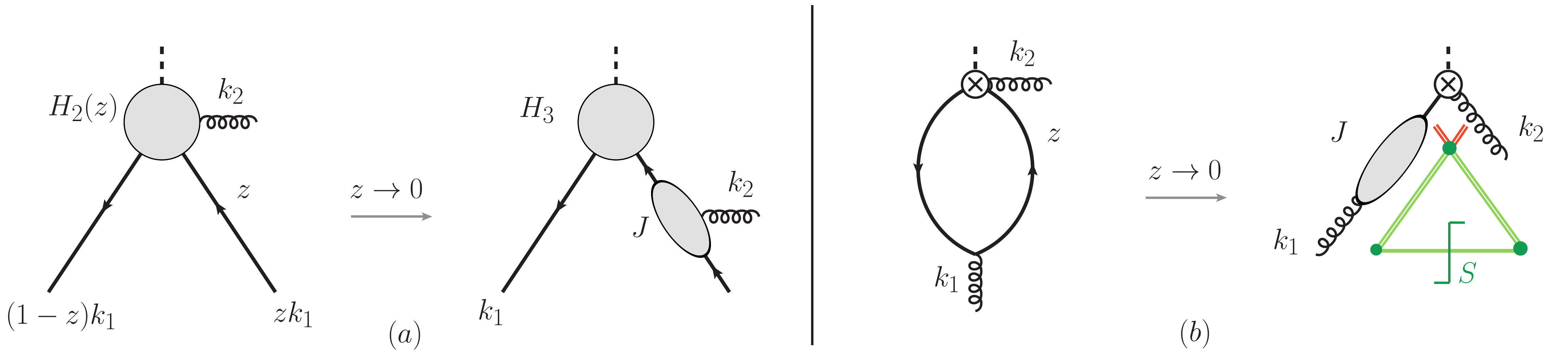}
	\caption{Graphical illustration of the refactorization conditions connecting different objects in the $gg\to h$ factorization formula to all orders of $\alpha_s$. The left panel portrays the first equation in \eqref{eq:refac1}, while the right panel illustrates the second equation.}
	\label{fig:refac}
\end{figure}

Complications arise because the integrals in the second and third term in \eqref{eq:factorizationPhoton} are endpoint divergent in the regions where $z\to 0$, $z\to 1$, and $\ell_\pm\to\infty$. From a physical point of view, these regions are at the boundary where a collinear quark becomes soft or a soft quark becomes collinear, hinting that both divergent terms should have a closely related structure. This was shown rigorously in \cite{Liu:2019oav,Liu:2020tzd}, where two refactorization conditions were proven to hold to all orders of perturbation theory. They are
\begin{equation}\label{eq:refac1}
	\begin{aligned}
		\braces{\bar{H}_{2,\gamma}^{(0)}(z)} 
		&= - H_{3,\gamma}^{(0)}J_\gamma^{(0)}(zM_h^2)\,, \\
		\braces{\langle O_{2,\gamma}^{(0)}(z)\rangle}
		&= - \frac{\varepsilon_1^\perp(k_1)\cdot\varepsilon_2^\perp(k_2)}{2}\int_0^\infty\frac{\df\ell_+}{\ell_+}J_\gamma^{(0)}(-M_h\ell_+)S_\gamma^{(0)}(zM_h\ell_+)\,.
	\end{aligned}
\end{equation}
The function $\bar H_2(z)$ is defined via
\begin{equation}
    H_2(z) = \frac{\bar H_2(z)}{z(1-z)} \,, 
\end{equation}
and the symbol $\braces{\,\dots}$ denotes that one should only keep the leading terms in the $z\to 0$ limit. The arguments in the proof can also be applied to the corresponding functions in the gluon case, for which analogous refactorization conditions hold. The situation is portrayed for the gluon case in figure~\ref{fig:refac}.

\begin{figure}
	\centering
	\includegraphics[width=0.55\textwidth]{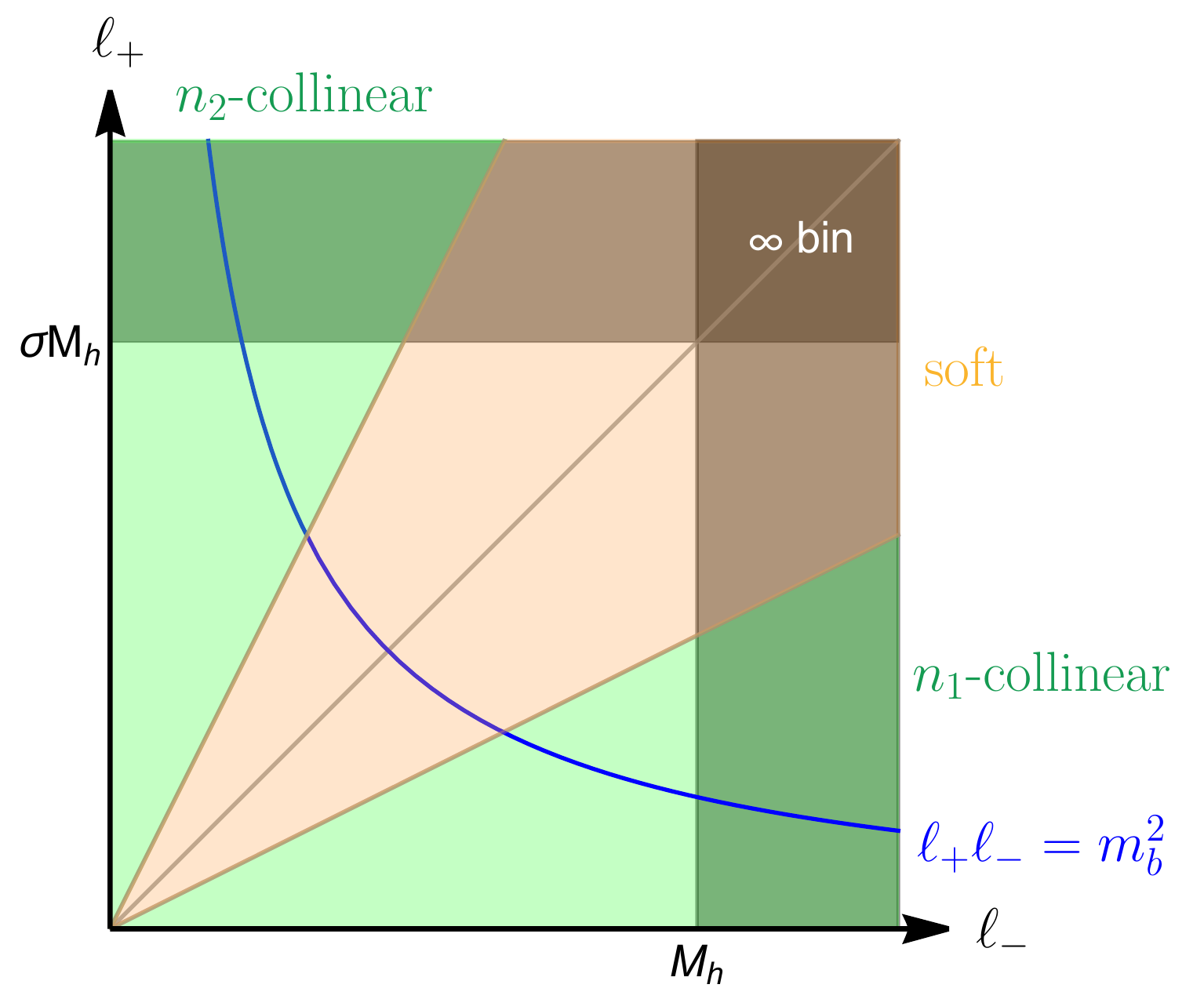}
	\caption{Graphical illustration of the impact of the cutoffs on the convolution integrals over $\ell_+$ and $\ell_-$ in the last term of the bare factorization formula \eqref{eq:facto2Photon}. The ``infinite bin'' is subtracted twice and must be added back in the form of an extra contribution to the bare Wilson coefficient $H_{1,\gamma}^{(0)}$.}
	\label{fig:infinitybin}
\end{figure}

Using these refactorization conditions allows one to rewrite the bare factorization theorem in a form that is free of endpoint divergences. The result is
\begin{equation}
	\label{eq:facto2Photon}
	\begin{aligned}
		&\mathcal{M}_b(h\to\gamma\gamma)=\left(H_{1,\gamma}^{(0)}+\Delta H_{1,\gamma}^{(0)}\right)\langle O_{1,\gamma}^{(0)}\rangle\\
		&+2\int_0^1\!\df z\bigg[H_{2,\gamma}^{(0)}(z)\langle O_{2,\gamma}^{(0)}(z)\rangle-\frac{\braces{\bar{H}_2^{(0)}(z)}}{z}\braces{\langle O_{2,\gamma}^{(0)}(z)\rangle}-\frac{\braces{\bar{H}_2^{(0)}(1-z)}}{1-z}\braces{\langle O_{2,\gamma}^{(0)}(z)\rangle}\bigg]\\[-2mm]
		&+\varepsilon_1^\perp\!\cdot\!\varepsilon_2^\perp\lim\limits_{\sigma\to-1}\!H_{3,\gamma}^{(0)}\!\int_0^{M_h}\!\frac{\df\ell_-}{\ell_-}\int_0^{\sigma M_h}\!\frac{\df\ell_+}{\ell_+}J_\gamma^{(0)}(M_h\ell_-)J_\gamma^{(0)}(-M_h\ell_+)S_\gamma^{(0)}(\ell_-\ell_+)\Big|_\text{leading power}\,.
	\end{aligned}
\end{equation}
Removing the divergences in the second term by a plus-type subtraction and applying the refactorization conditions introduces cutoffs on the integrals in the third term. Since, as shown in figure~\ref{fig:infinitybin}, the region $|\ell_\pm|\geq M_h$ is subtracted twice, this purely hard ``infinity-bin'' contribution must be added back, giving rise to the quantity $\Delta H_{1,\gamma}^{(0)}$ in \eqref{eq:facto2Photon}.

Renormalizing the quark mass $m_b$, the Yukawa coupling $y_b$ and the strong coupling $\alpha_s$ is not sufficient to remove all ultraviolet (UV) divergences from the bare operators and hard matching coefficients. The remaining divergences must be eliminated by renormalizing these objects themselves. This is in general a non-trivial task, since the renormalization factors must be applied in the convolution sense, and moreover the operators $O_{1,\gamma}$ and $O_{2,\gamma}$ mix under renormalization. Endpoint divergences in the renormalized factorization theorem are eliminated similarly to the bare case. An additional complication arises from the fact that, due to the presence of the cutoffs on the convolution integrals, the operations of renormalization and the removal of endpoint divergences do not commute. This leads to the appearance of so-called ``mismatch term'' \cite{Liu:2020tzd} that emerge when rearranging the expressions into the form of \eqref{eq:facto2Photon}. Since these mismatch terms only receive contributions from momentum regions above the Higgs mass scale, they can be collected into an additional contribution to the renormalized Wilson coefficient $H_{1,\gamma}(\mu)$. It is thus possible to derive a renormalized version of the factorization formula \eqref{eq:facto2Photon}. 

\subsection[Factorization theorem for $gg\to h$]{\boldmath Factorization theorem for $gg\to h$}
\label{subsec:factoGluon}

\begin{figure}[hbt]
	\centering
	\includegraphics[width=7cm]{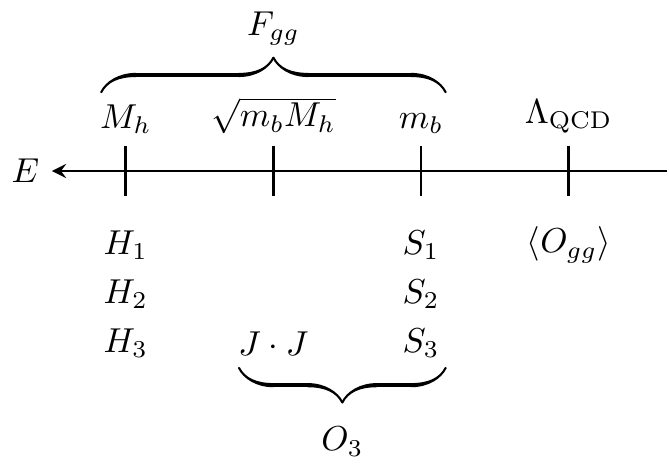}
	\caption{Illustration of the four energy scales relevant to the $gg\to h$ fusion process mediated via light quarks. The different objects in the factorization theorem are shown at their respective scales. The hard, jet and soft functions can be collected into the $h\to gg$ form factor $F_{gg}$. This quantity is the Wilson coefficient arising when the SM is matched onto a low-energy effective theory below the scale $m_b$.}\label{fig:skalen}
\end{figure}

Our goal in this work is to apply the methodology introduced above to the $gg\to h$ process, which is structurally very similar to the photon case, with the crucial difference that the external gluons carry color and are not infrared-safe asymptotic states. In fact, deriving the factorization theorem in the gluon case is a four-scale problem. The involved scales are the mass of the Higgs boson $M_h$, the mass of the light quark $m_b$, an intermediate scale $\sqrt{M_h m_b}$ only present for the analog of the third term in \eqref{eq:facto2Photon}, and the scale $\Lambda_\text{QCD}$, where non-perturbative effects come into play, accounting for the fact that the gluons are confined inside the colliding protons. The different scales and the corresponding objects in the factorization theorem are shown in figure~\ref{fig:skalen}. To deal with this situation, we consider the three-step matching procedure $\text{SM}\to\text{SCET}_1\to\text{SCET}_2\to\text{LEFT}$, where LEFT is the low-energy effective theory below the $b$-quark mass scale. In analogy with the photon case studied in \cite{Liu:2019oav,Liu:2020wbn}, the relevant SCET$_1$ operators are 
\begin{equation}\label{Ondef}
\begin{aligned}
   O_1 
   &= \frac{m_b}{g_s^2}\,h\,\G_{n_1}^{\perp\mu,a}\,\G_{n_2\,\mu}^{\perp a} \,, \\
   O_2(z) 
   &= h\,\Big[ \bar\X_{n_1}\gamma_\perp^\mu\spac T^a\spac\frac{\nbsl_1}{2}\,
    \delta(z\spac\bar n_1\!\cdot k_1+i\bar n_1\!\cdot\partial)\,\X_{n_1} \Big]\,
    \G_{n_2\,\mu}^{\perp,a} \,, \\
   O_3
   &= T\,\Big\{ h\,\bar\X_{n_1} \X_{n_2},
    i\!\int\!\!d^Dx\,{\cal L}_{q\,\xi_{n_1}}^{(1/2)}(x), 
    i\!\int\!\!d^Dy\,{\cal L}_{\xi_{n_2} q}^{(1/2)}(y) \Big\} 
    + \mbox{h.c.} \,,
\end{aligned}
\end{equation}
where $h$ denotes the Higgs field. Here and below, fields without an argument are located at the spacetime point $x=0$. The symbols $\G_{n_i}^{\mu,a}$ and $\X_{n_i}$ denote $n_i$-hard-collinear gluon and $b$-quark fields defined in SCET$_1$ (the so-called ``gauge-invariant building blocks'' \cite{Bauer:2002nz,Hill:2002vw}), which differ from the ordinary quantum fields $G^{\mu,a}$ and $\psi$ in that they contain hard-collinear Wilson lines in their definition and that they obey the constraints $\bar n_i\cdot\G_{n_i}^a=0$ and $\nsl_i\spac\X_{n_i}=0$. Note that the Feynman rule for the vector field $\G_{n_i}^{\mu,a}$ contains a factor of $g_s$, which is the reason why we have divided by $g_s^2$ in the definition of $O_1$. The operator $O_3$ contains the time-ordered product of the scalar Higgs current with two subleading-power terms in the SCET Lagrangian \cite{Beneke:2002ph}, in which hard-collinear fields are coupled to a soft quark field. When the above operators are matched into SCET$_2$, the hard-collinear fields in $O_1$ and $O_2$ are simply replaced by the corresponding collinear fields, whereas the operator $O_3$ is matched onto a double convolution of two jet functions with a soft function, as shown in \eqref{eq:O3factoPhoton} for the photon case. 

The only operator in the LEFT needed for our purposes is the two-gluon operator 
\begin{equation}
   O_{gg} = \frac{1}{g_s^2}\,\G_{n_1}^{\perp\mu,a}\,\G_{n_2\,\mu}^{\perp a} \,,
\end{equation}
built out of two collinear gluon fields along the directions $n_1$ and $n_2$. The matching relations for the relevant SCET operators onto the operator $O_{gg}$ involve soft functions $S_i$ as matching coefficients. For the case of Higgs-boson production at proton-proton colliders, we define
\begin{equation}
	\label{eq:matchO}
	\begin{aligned}
		\langle pp|\spac O_i\spac|h\rangle
		&= S_i\,\langle pp|\spac O_{gg}\spac|0\rangle \,; \quad i=1,2 \,,\\
		\langle pp|\spac O_3\spac|h\rangle 
		&= J\otimes J\otimes S_3\,\langle pp|\spac O_{gg}\spac|0\rangle \,.
	\end{aligned}
\end{equation}
Being Wilson coefficients, the soft functions $S_i$ can be calculated in perturbation theory using on-shell gluon states. 
All non-perturbative physics is incorporated in the matrix element $\langle pp|\spac O_{gg}\spac|0\rangle$. The operator $O_{gg}$ requires renormalization and hence its matrix elements are scale dependent. When the $gg\to h$ production amplitude is squared and integrated over phase space, the squared matrix element of $O_{gg}$ yields the product of two gluon distribution functions of the proton. 

\begin{figure}
	\centering
	\includegraphics[width=\textwidth]{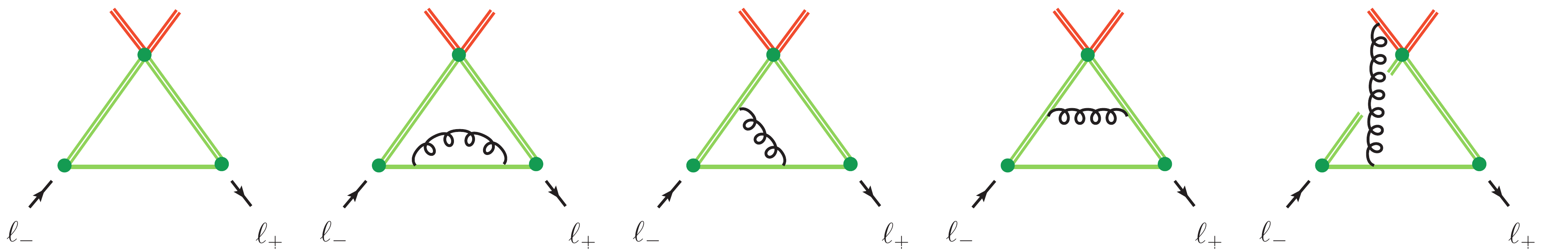}
	\caption{Feynman diagrams contributing to the calculation of the soft function $S_3$. We omit the mirror graphs of the third and last diagram.}
	\label{fig:soft}
\end{figure}

The hard, jet and soft functions can be combined into a perturbatively calculable short-distance quantity referred to as the $gg\to h$ form factor $F_{gg}$. The interpretation of the total matrix element as a product of a form factor and a non-perturbative low-energy gluon matrix element allows for the identification of the form factor as the non-abelian counterpart of the $h\to\gamma\gamma$ amplitude. The calculation of the hard, jet and soft functions then proceeds in an analogous way as in the photon case. Following the arguments presented above, we write the bare factorization theorem for the light-quark induced contribution to the $gg\to h$ form factor in the form
\begin{equation}
	\label{eq:factoGluon}
		\begin{aligned}
		F_{gg}^{(0)}&=\overbrace{\left(H_1^{(0)}+\Delta H_1^{(0)}\right)S_1}^{T_1^{(0)}} +\overbrace{4\int_0^1\frac{\df z}{z}\Big(\bar{H}_2^{(0)}(z)S_2^{(0)}(z) -\braces{\bar{H}_2^{(0)}(z)}\braces{S_2^{(0)}(z)}\Big)}^{T_2^{(0)}}\\
		&~~ +\underbrace{\lim_{\sigma\to -1}H_3^{(0)}\!\int_0^{M_h}\!\frac{\df\ell_-}{\ell_-}\int_0^{\sigma M_h}\!\frac{\df \ell_+}{\ell_+}J^{(0)}(-M_h\ell_-)J^{(0)}(M_h\ell_+)S_3^{(0)}(\ell_-\ell_+)\Big|_{\text{leading power}}}_{T_3^{(0)}} \,,
	\end{aligned}
\end{equation}
which is free of endpoint divergences and UV finite. Note that due to the cutoffs the third term contains some power-suppressed contributions, which should be dropped for consistency. The remaining infrared (IR) poles will eventually be absorbed by the renormalization of the operator $O_{gg}$. The fact that the integrand of the second term is symmetric under exchange $z\leftrightarrow(1-z)$ explains the additional factor 2 in front of the integral compared with \eqref{eq:facto2Photon}. The bare hard coefficients $H_i^{(0)}$ and soft functions $S_1^{(0)}$ and $S_2^{(0)}$ are defined and calculated in analogy with the photon case. The corresponding expressions can be found in appendix~\ref{app:BareQuant}. The jet function for the gluon case has been calculated at two-loop order in \cite{Liu:2021mac}. An important difference with respect to the photon case concerns the soft function $S_3^{(0)}$, which is related to the structure
\begin{equation}
	\label{eq:defS3}
			W_{ab}^{\alpha\beta}(x_-, y_+)=\mathrm{\hat{T}}\,\mbox{Tr}_c \left[
			S_{n_2}(0)T^bS^\dagger_{n_2}(y_+)\,q_s^\beta(y_+)\,
	\bar q_s^\alpha(x_-)\,S_{n_1}(x_-)T^aS^\dagger_{n_1}(0)\right] .
\end{equation} 
Here Tr$_c$ indicates a trace over color indices and $\hat{\text{T}}$ stands for time ordering. $S_{n_i}$ denotes a soft Wilson line in the direction $n_i$. The position variables are defined as $x_-^\mu=\frac{n_1^\mu}{2}\,(n_2\cdot x)$ and $y_+^\mu=\frac{n_2^\mu}{2}\,(n_1\cdot x)$. In contrast to the photon case, it is not possible to combine the semi-finite soft Wilson lines $S_{n_2}(0)$ and $S^\dagger_{n_2}(y_+)$ into a finite-length Wilson line because of the insertion of the  color generator $T^b$, and similarly for the soft Wilson lines in the $n_1$ direction. We may however use the identity
\begin{equation}
	\label{eq:Wilsonlinesrelation}
	S_{n_i}(x)T^aS^\dagger_{n_i}(x)=(\mathcal{Y}_{n_i}(x))^a_{\\,,b}~ T^b\,,
\end{equation}
with $\mathcal{Y}_{n_i}(x)$ a semi-finite soft Wilson line in the adjoint representation, to obtain
\begin{equation}
	\label{eq:softopadj}
	\begin{aligned}
		W_{ab}^{\alpha\beta}(x_-, y_+)=\mathrm{\hat{T}}\,\mbox{Tr}_c \left[ \mathcal{Y}_{n_2}(0)^b_{\,d}\spac T^d\,S_{n_2}(0,y_+)\,q_s^\beta(y_+)\,
		\bar q_s^\alpha(x_-)\,S_{n_1}(x_-,0)\,\mathcal{Y}_{n_1}(0)^a_{\,c}\spac T^c\right] ,
	\end{aligned}
\end{equation} 
with 
\begin{equation}
	\label{eq:Wilsonsegment}
	\begin{aligned}
		S_{n_2}(0,y_+)\equiv S_{n_2}(0)S_{n_2}^\dagger(y_+)=\hat{\mathrm{P}}\exp\left[ig_s\int_{y_+}^0\df t\, n_2\cdot G_s^a(tn_2)\, T^a \right] .
	\end{aligned}
\end{equation}
Here, $G_s^a(x)$ is a soft gluon field without any Wilson line dressing. The Feynman diagrams contributing to the correlator $W_{ab}^{\alpha\beta}$ up to two-loop order are shown in Figure~\ref{fig:soft}. They consist of ``tipi-tent'' graphs, in which we represent the finite-length Wilson lines in the fundamental representation as green double lines, whereas the semi-finite Wilson lines in the adjoint representation are drawn as red double lines. The last diagram, in which the gluon connects to one of the semi-finite Wilson lines, is absent in the photon case considered in \cite{Liu:2019oav,Liu:2020eqe}. The soft function $S_3^{(0)}$ is defined in terms of the discontinuity of $W_{ab}^{\alpha\beta}$ in momentum space. 

\subsection[Bare expression for the $gg\to h$ form factor]{\boldmath Bare expression for the $gg\to h$ form factor}

To show that all UV divergences cancel in the sum of the three terms in \eqref{eq:factoGluon}, we first express the bare parameters, i.e.\ the $b$-quark mass, the $b$-quark Yukawa coupling and the strong coupling $\alpha_s$, in terms of renormalized parameters. The relevant renormalization conditions are given in appendix~\ref{app:RenFac}. We use the running parameters $m_b(\mu)$ and $y_b(\mu)$ in the overall prefactor of the form factor. However, in the arguments of logarithms we use the $b$-quark pole mass $m_b$. Since the form factor is calculated using on-shell gluon states, it is IR divergent. We remove the IR poles by multiplying with the renormalization factor $Z_{gg}^{-1}$, where $Z_{gg}$ is the UV renormalization factor of the two-gluon operator $O_{gg}$, defined by $O_{gg}(\mu)=Z_{gg}\spac O_{gg}^{(0)}$. In the $\overline{\text{MS}}$-scheme, it is given by \cite{Becher:2009cu}
\begin{equation}
	\label{eq:Zggdef}
Z_{gg}=1-\frac{\alpha_s(\mu)}{4\pi}\left[ 2 C_A \left( \frac{1}{\epsilon^2} - \frac{L_h}{\epsilon} \right)
+\frac{\beta_0}{\epsilon}\right]+\mathcal{O}(\alpha_s^2)\,,
\end{equation}
where $L_h=\ln[(-M_{h}^2-i0)/\mu^2]$. We write the result for the $gg\to h$ form factor as
\begin{equation}
	\label{eq:formfactor}
	Z_{gg}^{-1} F_{gg}^{(0)} =\mathcal{M}_0\,Z_{gg}^{-1} \left( T_1^{(0)}+T_2^{(0)}+T_3^{(0)} \right) ,
\end{equation}
with the overall prefactor
\begin{equation}
	\label{eq:M0}
	\mathcal{M}_0=T_F\spac\delta_{ab}\spac\frac{\alpha_{s}(\mu)}{\pi}\spac\frac{y_{b}(\mu)}{\sqrt{2}}\spac m_b(\mu) \,.
\end{equation} 
The three contributions read
\begin{align}\label{eq:T1T2T3bare}
	Z_{gg}^{-1}\spac T_1^{(0)} &= -2+\frac{\alpha_s(\mu)}{4\pi}\Bigg\{C_F\bigg[-\frac{\pi ^2}{3 \epsilon ^2}+\frac{1}{\epsilon }\left(\frac{2 \pi ^2 L_h}{3}-10 \zeta _3\right)-\frac{2 \pi ^2}{3} L_h^2+4 \left(5 \zeta _3+3\right) L_h\notag\\
	&\quad -36-\frac{7 \pi ^4}{30}\bigg]+C_A\bigg[\frac{\pi^2}{3\epsilon^2}-\frac{1}{\epsilon}\left(\frac{2 \pi ^2 L_h}{3}-10 \zeta _3\right)+\left(2+\frac{2\pi ^2}{3}\right) L_h^2-20 \zeta _3 L_h\notag\\
	&\quad -12-\frac{\pi ^2}{6}+18\zeta_3+\frac{ \pi ^4}{5}\bigg]\Bigg\} + \mathcal{O}(\alpha_s^2) \,,\notag\\
	Z_{gg}^{-1}\spac T_2^{(0)} &= \frac{\alpha_s(\mu)}{4\pi}\Bigg\{C_F\bigg[\frac{\pi ^2}{3 \epsilon ^2}+\frac{1}{\epsilon }\bigg(2 \zeta _3-\frac{2 \pi ^2 L_h}{3}\bigg)+\frac{\pi ^2}{3}  \big(L_h^2-L_m^2\big)+L_h \left(\frac{2 \pi ^2 L_m}{3}-4 \zeta _3\right)\notag\\
	&\quad +8 \zeta _3+\frac{13\pi ^4}{90}\bigg]+C_A\bigg[-\frac{\pi ^2}{3 \epsilon ^2}+\frac{1}{\epsilon }\bigg(\frac{2 \pi ^2 L_h}{3}-6 \zeta _3\bigg)-\frac{\pi ^2}{3}  \big(L_h^2-L_m^2\big)\notag\\
	&\quad +L_h \left(4 \zeta _3-\frac{2 \pi ^2 L_m}{3}\right)+8 \zeta _3 L_m-\frac{\pi ^2}{6}-6 \zeta _3-\frac{\pi ^4}{45}\bigg]\Bigg\} + \mathcal{O}(\alpha_s^2) \,,\notag\\
	Z_{gg}^{-1}\spac T_3^{(0)} &= \frac{L^2}{2}+\frac{\alpha_s(\mu)}{4\pi}\Bigg\{C_F\bigg[\frac{8 \zeta _3}{\epsilon }-\frac{L^4}{12}-L^3+L^2 \left(-3 L_m-\frac{\pi ^2}{3}+4\right) \notag\\
	&\quad + \bigg(16-12L_m+\frac{2 \pi ^2 }{3}\bigg)L-16 \zeta _3 L_m-4 \zeta _3-\frac{\pi ^4}{9} \bigg]\notag\\
	&\quad +C_A\bigg[-\frac{4\zeta_3}{\epsilon}-\frac{5 L^4}{12}-L^3L_m-\frac{L^2L_m^2}{2}+\left(1+\frac{\pi^2}{12}\right)L^2+4\zeta_3(L+2L_m)\bigg] \Bigg\} \notag\\
	&\quad + \mathcal{O}(\alpha_s^2) \,.
\end{align}
The different logarithms appearing in the expressions are
\begin{equation}
	L_h = \ln\frac{-M_h^2-i 0}{\mu^2} \,, \qquad
	L_m = \ln\frac{m_b^2}{\mu^2} \,, \qquad
	L = L_h - L_m = \ln\frac{-M_h^2-i 0}{m_b^2} \,,
\end{equation}
with $m_b$ being the pole mass. It can readily be checked that the remaining $1/\epsilon$ poles cancel in the sum of the three contributions. Consequently, we find for the full form factor
\begin{align}
	\label{eq:ampfrombare}
		Z_{gg}^{-1} F_{gg}^{(0)}
		&= \mathcal{M}_0\Bigg\{-2+\frac{L^2}{2}+\frac{\alpha_s(\mu)}{4\pi}\Bigg[C_A\Bigg(-\frac{5L^4}{12}-L^3L_m-\frac{L^2L_m^2}{2}+\left(3+\frac{5\pi^2}{12}\right)L^2 \notag\\
		&\qquad +4LL_m+2L_m^2-12\zeta_3L-12-\frac{\pi^2}{3}+12\zeta_3+\frac{8\pi^4}{45} \Bigg) \notag\\
		&\quad +C_F\Bigg(-\frac{L^4}{12}-L^3-3L_mL^2 +\left(4-\frac{2\pi^2}{3}\right)L^2
		 +\left(16 \zeta _3+\frac{2 \pi ^2}{3}+12\right)L \notag\\
		&\qquad +12L_m-36+4 \zeta _3-\frac{\pi ^4}{5} \Bigg) \Bigg] + \mathcal{O}(\alpha_s^2) \Bigg\}\,.
\end{align} 
This result agrees with a corresponding expression obtained in \cite{Aglietti:2006tp} after taking into account some differences in the IR subtraction schemes. In the limit $C_A\to0$, and performing some simple replacements in the prefactor $\mathcal{M}_0$, the above result reproduces the two-loop amplitude for $h\to\gamma\gamma$ decay obtained in \cite{Liu:2019oav}.

\section{Renormalized factorization formula}
\label{sec:reno}

In this section, we establish the factorization formula in terms of renormalized quantities, which reads
\begin{equation}
	\label{eq:refact}
	\begin{aligned}
		F_{gg}(\mu) =\overbrace{H_1(\mu)S_1(\mu)}^{T_1(\mu)} +\overbrace{4\int_0^1\frac{\df z}{z}\Big(\bar{H}_2(z,\mu)S_2(z,\mu) -\braces{\bar{H}_2(z,\mu)}\braces{S_2(z,\mu)}\Big)}^{T_2(\mu)}\\
		+\underbrace{\lim_{\sigma\to -1}H_3(\mu)\int_0^{M_h}\frac{\df\ell_-}{\ell_-}\int_0^{\sigma M_h}\frac{\df\ell_+}{\ell_+}J(M_h\ell_-,\mu)J(-M_h\ell_+,\mu)S_3(\ell_-\ell_+,\mu)\Big|_{\text{leading power}}}_{T_3(\mu)} \,.
	\end{aligned}
\end{equation}
In general, we obtain the renormalized operators $O_i(\mu)$ from the bare operators $O_j^{(0)}$ using the relation
\begin{equation}\label{Zijdef}
   O_i(\mu) = Z_{ij}\spac O_j^{(0)} \,,
\end{equation}
where in some cases the product must be replaced by a convolution. The hard matching coefficients are renormalized with the inverse renormalization factors. As discussed in great detail in \cite{Liu:2020wbn}, to derive such a renormalized factorization theorem from the bare one is a non-trivial task. In addition to renormalizing the various ingredients, one needs to assure that renormalization does not reintroduce endpoint divergences. It can be shown that while renormalization and the subtraction of endpoint divergences do not commute, moving from the bare to the renormalized factorization theorem only introduces additional finite terms, which only depend on the hard scale $M_h$. These terms can hence be absorbed into a redefinition of the renormalized hard matching coefficient $H_1(\mu)$. 

\subsection[Renormalization of $T_3$]{\boldmath Renormalization of $T_3$}
\label{sec:T3renorm}

The hard function $H_3$ is the same as in the photon case $H_{3,\gamma\gamma}$, and so we can directly quote the corresponding expression from \cite{Liu:2020wbn}, which reads
\begin{equation}
	\label{eq:H3re}
	H_{3}(\mu)=Z_{33}^{-1} H_3^{(0)}=\frac{y_{b}(\mu)}{\sqrt{2}}\left[-1+\frac{C_{F} \alpha_{s}}{4 \pi}\left(L_{h}^{2}+2-\frac{\pi^{2}}{6}\right)\right] +\mathcal{O}(\alpha_s^2) \,.
\end{equation}
We collect all renormalization factors in appendix~\ref{app:RenFac} unless stated otherwise. Note that here and in the following we will suppress the scale dependence of the strong coupling constant and denote $\alpha_s\equiv \alpha_s(\mu)$ in the $\overline{\text{MS}}$-scheme with $n_f=5$ active quark flavors.

The radiative jet function is renormalized in the convolutional sense, i.e.\
\begin{equation}
	\label{eq:JetRedef}
	J(p^2,\mu)=\int_0^\infty\df x\,Z_J(p^2,xp^2)J^{(0)}(xp^2)\,.
\end{equation}
Both the bare function $J^{(0)}$ and the renormalized function $J$ have been calculated at two-loop order in \cite{Liu:2021mac}. One finds
\begin{equation}
	\label{eq:JetRe}
	J(p^2,\mu)=1+\frac{\alpha_s}{4\pi}(C_F-C_A)\left[L_p^2-1-\frac{\pi^2}{6}\right]+\mathcal{O}(\alpha_s^2) \,.
\end{equation}

Also the soft function is renormalized by means of a convolution, such that
\begin{equation}
	\label{eq:Softredef}
	S_3(w,\mu)=\int_0^\infty \df w^\prime \,Z_S(w,w^\prime)S_3^{(0)}(w^\prime)\,.
\end{equation}
In the photon case, the form of the renormalization factor $Z_S$ was deduced by applying RG consistency arguments to $T_3|_{h\to\gamma\gamma}$ \cite{Liu:2020eqe}. Later, Bodwin {\it et al.} have verified this conjecture by an explicit calculation \cite{Bodwin:2021cpx}. Following the same approach as in \cite{Liu:2020eqe}, we find the renormalization factor of the soft function in the gluon case to be
\begin{align}
	\label{eq:ZS}
		Z_S(w,w^\prime)&=\frac{w}{w^{\prime}}\,Z_{gg}^{-1} Z_{33} \int_{0}^{\infty} \frac{\df x}{x} Z_{J}^{-1}\left(\frac{M_{h} w^{\prime}}{x \ell_{+}}, \frac{M_{h} w}{\ell_{+}}\right) Z_{J}^{-1}\left(-x M_{h} \ell_{+},-M_{h} \ell_{+}\right)\notag\\
		&=\delta(w-w^\prime)+\frac{\alpha_s}{2\pi}\Bigg\{\left[(C_F-C_A)\bigg(\frac{1}{\epsilon^2}-\frac{ L_w}{\epsilon}\bigg)-\frac{3C_F-\beta_0}{2\epsilon}\right]\delta(w-w^\prime)\notag\\
		&\hspace{3.7cm} -\frac{2C_F-C_A}{\epsilon}\spac w\spac\Gamma(w,w^\prime)\Bigg\} + \mathcal{O}(\alpha_s^2) \,,
\end{align}
with $L_w=\ln(w/\mu^2)$. Here
\begin{equation}
	\label{eq:LN}
	\Gamma(y,x)=\left[\frac{\theta(x-y)}{x(x-y)}+\frac{\theta(y-x)}{y(y-x)}\right]_+
\end{equation}
is the Lange-Neubert kernel \cite{Lange:2003ff}. Note that the color factor in front of this distribution is $(2C_F-C_A)$, which differs from the color factor in front of the cusp logarithm $L_m$. We will see that this significantly complicates the solution of the RG evolution equation for the soft function compared with the photon case. The Lange-Neubert kernel plays a crucial role already at order $\mathcal{O}(\alpha_s)$, because the leading-order soft function is not a constant. Using \eqref{eq:Softredef} and \eqref{eq:ZS}, we find
\begin{equation}
	\label{eq:softrenormalized}
	\begin{aligned}
		S_3(w,\mu)=-\frac{T_F\spac\delta_{ab}\,\alpha_s}{\pi}\spac m_{b}(\mu)\,\Big[
		S_a(w,\mu)\theta\big(w-m_b^2\big)+S_b(w,\mu)\theta\big(m_b^2-w\big)\Big] \,,
	\end{aligned}
\end{equation}
with
\begin{equation}
\begin{aligned}
\label{eq:softrenormalizedcont}
	S_a(w,\mu) 
	&= 1 + \frac{\alpha_s}{4\pi}\spac\Bigg\{C_F\bigg[-L_w^2-6L_w+12-\frac{\pi^2}{2}
	 +2\operatorname{Li}_2\left(\hat{w}^{-1}\right) \\
	&\qquad -4\ln\left(1-\hat{w}^{-1}\right) \bigg(\frac{3}{2}\ln \hat{w}
	 +\ln\left(1-\hat{w}^{-1}\right)+L_m+1\bigg)\bigg] \\
	&\quad +C_A\bigg[L_w^2-\frac{\pi^2}{6}+2\operatorname{Li}_2\left(\hat{w}^{-1}\right) \\
	&\qquad +2\ln\left(1-\hat{w}^{-1}\right)\bigg(\ln\hat{w}+\ln\left(1-\hat{w}^{-1}\right)+L_m\bigg)\bigg]\Bigg\} 
	 + \mathcal{O}(\alpha_s^2) \,,\\
	S_b(w,\mu)&=\frac{\alpha_s}{4\pi}\left(C_F-\frac{C_A}{2}\right)4\ln(1-\hat{w})\big(\ln(1-\hat{w})+L_m\big) 
	 + \mathcal{O}(\alpha_s^2) \,,
\end{aligned}
\end{equation}
with $\hat{w}=w/m_b^2$. 

\subsection[Renormalization of $T_2$]{\boldmath Renormalization of $T_2$}
\label{sec:T2renorm}

The hard function $H_2$ is renormalized multiplicatively in the convolution sense. We find
\begin{equation}
    \label{eq:H2renormalized}
    \begin{aligned}
        \bar{H}_2(z,\mu) &= \int_0^1\df z^\prime\,Z_{22}^{-1}(z,z^\prime) \bar{H}_2^{(0)}(z^\prime)\\
		&=1+\frac{\alpha_s}{4\pi}\bigg\{C_F\bigg[ 2L_h\big(L_z+L_{\bar{z}}\big)+L_z^2+L_{\bar{z}}^2-3
		\bigg]\\
		&\quad +C_A\bigg[-L_h^2-2L_h\big(L_z+L_{\bar{z}}\big)-L_z^2-L_{\bar{z}}^2+1+\frac{\pi^2}{6}\bigg]\bigg\}
		 + \mathcal{O}(\alpha_s^2) \,,	
    \end{aligned}
\end{equation}  
and
\begin{align}
\label{eq:H2barrenormalized}
		\braces{\bar{H}_2(z,\mu)} &= \int_0^\infty\df z^\prime \,\braces{Z_{22}^{-1}(z,z^\prime)}\braces{\bar{H}_2^{(0)}(z^\prime)} \notag\\
		&=1+\frac{\alpha_s}{4\pi}\left[C_F\left(2L_h L_z+L_z^2-3\right)+C_A\left(-L_h^2-2L_h L_z-L_z^2+1+\frac{\pi^2}{6}\right)\right] \notag\\
		&\quad + \mathcal{O}(\alpha_s^2) \,,
\end{align}
where once again the renormalization factors can be found in appendix~\ref{app:RenFac}. Writing the evolution equation for the function $\bar{H}_2(z,\mu)$ instead of $H_2(z,\mu)$ changes the renormalization factor from $Z_{22}^{-1}(z',z)$ to
\begin{equation}
   \frac{z}{z'}\,Z_{22}^{-1}(z',z) = Z_{22}^{-1}(z,z') \,,
\end{equation}
which leads to the form shown above. To keep the expressions compact we have abbreviated $L_z=\ln z$ and $L_{\bar{z}}=\ln(1-z)$. Note that the result for $\braces{\bar{H}_2(z,\mu)}$ can also be obtained by keeping only the leading terms in the $z\to0$ limit in $\bar{H}_2(z,\mu)$.

The full form factor must be multiplied with an additional renormalization factor $Z_{gg}^{-1}$. Therefore in the renormalization condition for the soft function $S_2$ this factor also appears. Additionally, $Z_{22}^{-1}$ depends on the hard scale $M_h$ via the logarithm $L_h$, but the soft function should only depend on the soft scale $m_b$. This is indeed the case when we combine the two renormalization factors. Furthermore, in analogy with the photon case we find that $S_1$ and $S_2$ mix under renormalization. Hence, the renormalization condition takes the form
\begin{equation}
	\label{eq:O2generalrenorm}
	\begin{aligned}
		S_2(z,\mu) = Z_{gg}^{-1} \left[ \int_0^1\df z^\prime Z_{22}(z,z^\prime)\spac S_2^{(0)}(z^\prime)
		+Z_{21}(z)\spac S_1^{(0)} \right] .
	\end{aligned}
\end{equation}
For the renormalized soft function, we then obtain (with $\bar z\equiv 1-z$)
\begin{align}
	\label{eq:O2renormalized}
		 S_2(z,\mu) &= \frac{T_F\spac\delta_{ab}\alpha_s}{2\pi}\spac m_b(\mu)\,\Bigg\{-L_m +\frac{\alpha_s}{4\pi}\bigg[C_F\bigg(L_{m}^{2}\big(L_z+L_{\bar{z}}+3\big) \notag\\
		&\quad -L_{m}\bigg(L_z^2+L_{\bar{z}}^2-4 L_z  L_{\bar{z}}+11-\frac{2 \pi^{2}}{3}\bigg)+F(z)+F(\bar z)\bigg) \notag\\
		&\quad +C_A\bigg(-L_m^2\big(L_z+L_{\bar{z}}\big)+L_m\big(L_z^2+L_{\bar{z}}^2-1\big)+G(z)+G(\bar z)\bigg)\bigg]
		 + \mathcal{O}(\alpha_s^2) \Bigg\} \,,
\end{align}
with
\begin{equation}
	\label{eq:FG}
	\begin{aligned}
		F(z) &= \frac{L_z^3}{6}+L_z^2\big( z-L_{\bar{z}}\big)-L_z \left(-L_{\bar{z}}+\frac{1+3 z}{2}\right)-(4 L_z+2 z) \operatorname{Li}_{2}(z)\\
		&\quad +6 \operatorname{Li}_{3}(z)+\frac{11}{2}-4 \zeta_{3}\,,\\
		G(z) &= -\frac{L_z^3}{6}-\frac{z}{2}L_z^2+\frac{1}{2}\left(1+2z-L_{\bar{z}}\right)L_z+(2 L_z-(1-z))\operatorname{Li}_{2}(z)\\
		&\quad -4\operatorname{Li}_{3}(z)+\frac{1}{2}+4\zeta_3\,.
	\end{aligned}
\end{equation}

\subsection[Renormalization of $T_1$]{\boldmath Renormalization of $T_1$}
\label{subsec:T1ren}

\begin{figure}[t]
	\begin{center}
		\subfigure[The phase space of mismatch in $T_2$ (yellow) and $T_3$ (orange).]{
			\label{fig:mismatch1}\includegraphics[width=0.70\textwidth]{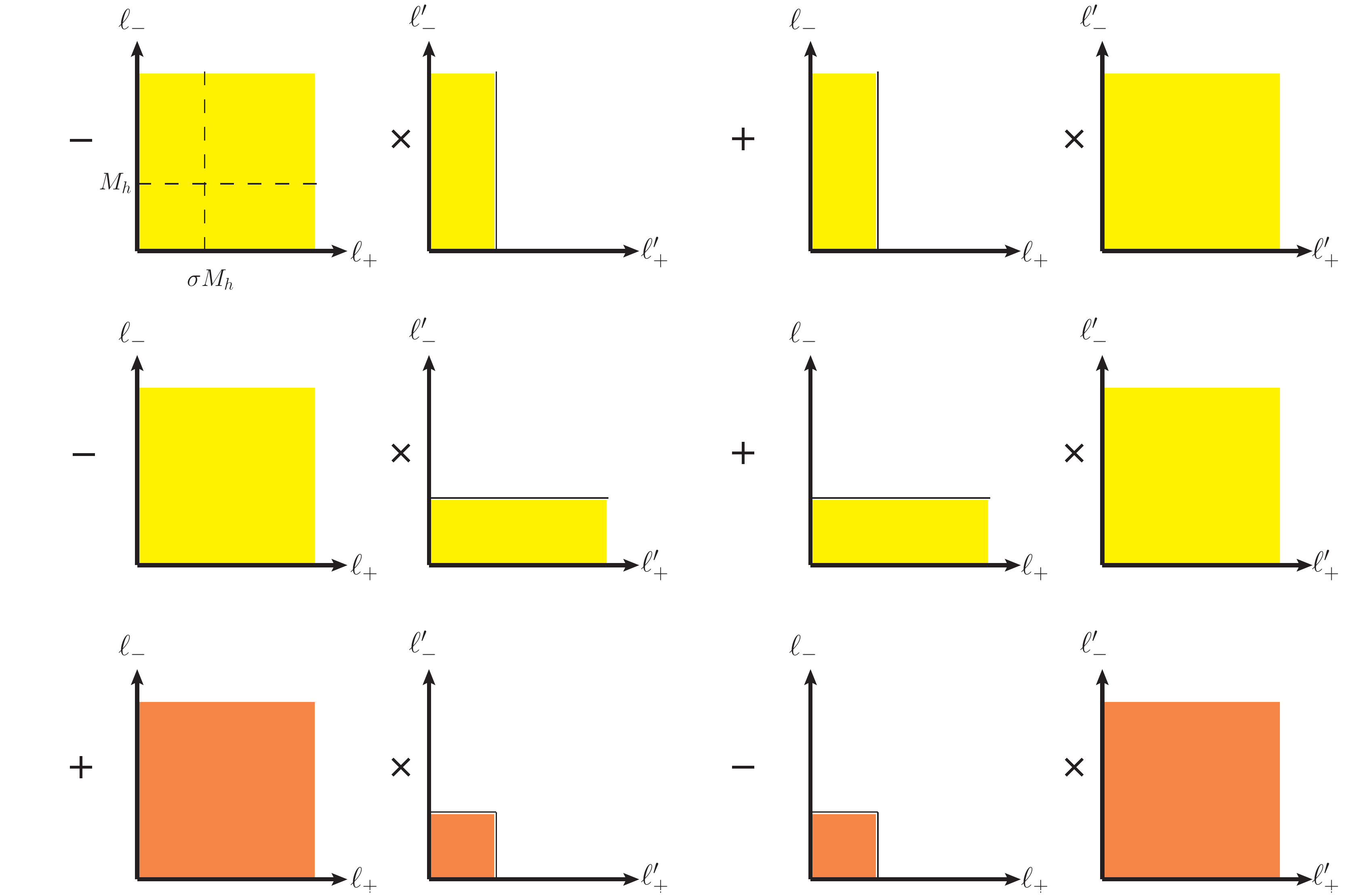}
		}\\
		\vspace{5mm}
		\subfigure[The combination of mismatch in $T_2$ and $T_3$ is given by phase-space integration in the purple region. It can be further flipped into the blue region, which is purely hard.]{
			\label{fig:mismatch2}\includegraphics[width=0.70\textwidth]{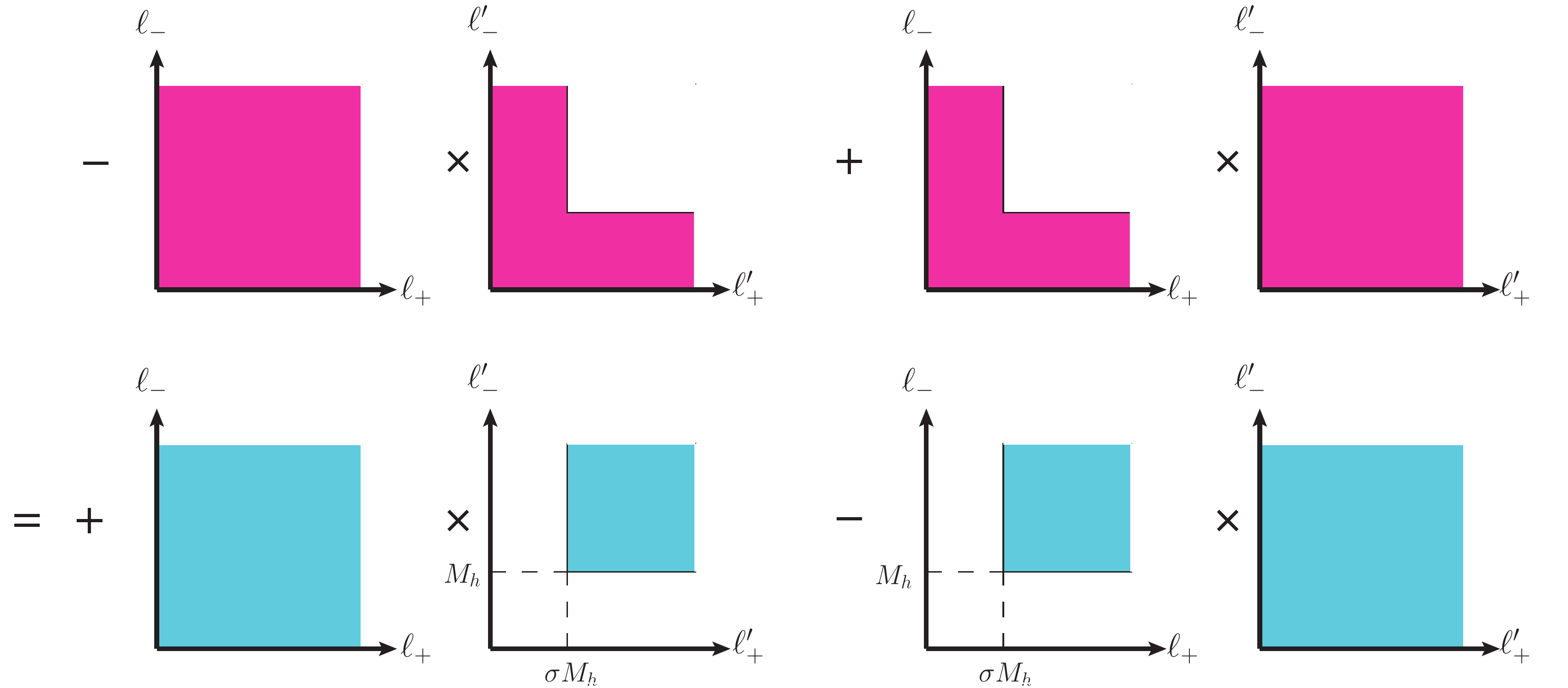} 
		} 
		\caption{\label{fig:mismatch}  The phase space of mismatch in $T_2$ and $T_3$.}
	\end{center}
\end{figure}

The renormalization condition for the hard function $H_1(\mu)$ is given by 
\begin{equation}\label{eq:H1defcont}
	\begin{aligned}
		H_1(\mu) &= Z_{11}^{-1} \left( H_1^{(0)} +\Delta H_1^{(0)}-\delta'\!H_1-\delta H_1 \right) \\
		&\quad +4\int_0^1\frac{\df z}{z}\Big(\bar{H}_2^{(0)}(z)Z_{21}^{-1}(z)-\braces{\bar{H}_2^{(0)}(z)}\braces{Z_{21}^{-1}(z)} \Big)\,,
	\end{aligned}
\end{equation}
where $\Delta H_1^{(0)}$ denotes the contribution from infinity-bin subtraction. Note that in this case the renormalization factor $Z_{gg}^{-1}$ must be associated with the hard matching contribution and not with the soft function $S_1$. The counterterms $\delta'\!H_1$ and $\delta H_1$ account for the ``mismatch contributions'' in $T_2$ and $T_3$, respectively \cite{Liu:2020wbn}. Using the relation between the renormalization factors in (\ref{eq:ZS}), $\Delta H_1^{(0)}$ can be written as 
\begin{equation}\label{eq:infinitybinH1}
	\begin{aligned}
		Z_{gg}^{-1}\spac \Delta H_1^{(0)} S_1^{(0)}
		&= - H_3^{(0)}\spac Z_{33}^{-1}
		 \int_{M_h}^\infty\!\df\ell_- \int_0^\infty\!\frac{\df\ell_-'}{\ell_-'}\,
		 \int_{\sigma M_h}^\infty\!\df\ell_+ \int_0^\infty\!\frac{\df\ell_+'}{\ell_+'} \\
		&\quad\times \int_0^\infty\df w\,S^{(0)}(w) J^{(0)}(-M_h\ell_+) J^{(0)}(M_h\ell_-) \\
		&\quad\times Z_J(M_h\ell_-',M_h\ell_-)\,Z_J(-M_h\ell_+', -M_h\ell_+) Z_S(\ell_+'\ell_-',w)\,.
	\end{aligned}
\end{equation}
Owing to the refactorization conditions for $\braces{\bar{H}_2^{(0)}(z)}$ and $\braces{S_2^{(0)}(z)}$ shown in \eqref{eq:refac1}, $\delta'\!H_1$ and $\delta H_1$ can be expressed in terms of four-fold integrals with the same integrand as in (\ref{eq:infinitybinH1}), but with different integration limits. The yellow and orange regions in figure~\ref{fig:mismatch} correspond to the integral domains relevant for $\delta'\!H_1$ and $\delta H_1$, respectively. Adding them up, the resulting integration in the purple region can be further flipped into the blue region, because the four-fold integration in the entire region is scaleless. In addition, the integration in the second blue region eliminates the contribution from $\Delta H_1$ exactly. As a result, the renormalized coefficient $H_1(\mu)$ can be expressed as
\begin{equation}\label{eq:H1ren}
	\begin{aligned}
		H_1(\mu) &= Z_{11}^{-1} H_1^{(0)} 
		  +4\int_0^1\frac{\df z}{z}\Big(\bar{H}_2^{(\epsilon)}(z,\mu)Z_{21}(z)-\braces{\bar{H}^{(\epsilon)}_2(z,\mu)}\braces{Z_{21}(z)} \Big)Z_{11}^{-1} \\
		&\quad -H_3(\mu)\lim_{\sigma\to -1} \int_{M_h}^{\infty}\frac{\df\ell_-}{\ell_-} \int_{\sigma M_h}^{\infty}\frac{\df\ell_+}{\ell_+} J^{(\epsilon)}(M_h\ell_-,\mu)J^{(\epsilon)}(-M_h\ell_+,\mu) 
		\spac\frac{S_3^{(\epsilon)}(\ell_+\ell_-,\mu)}{S_1(\mu)} \,,
		\end{aligned}
\end{equation}
where the superscripts ``$(\epsilon)$'' in $J$, $S$ and $\bar H_2$ indicate that the full dependence on the dimensional regulator must be kept in place after renormalization, as explained in \cite{Liu:2020wbn}. This form makes it explicit that $H_1(\mu)$ only depends on the hard scale $M_h$ to all orders in $\alpha_s$. The explicit result for this function at next-to-leading order (NLO) in perturbation theory is
\begin{align}
\label{eq:H1final}
		H_1(\mu)&=\frac{y_b(\mu)}{\sqrt{2}}\frac{T_F\spac\delta_{ab}\spac\alpha_s}{\pi}\Bigg\{\! 
		-2+\frac{\alpha_s}{4\pi}\bigg[C_F\bigg(\! -\frac{\pi^2}{3}L_h^2+(12+8\zeta_3)L_h-36-\frac{2\pi^2}{3}-\frac{11\pi^4}{45}\bigg)\notag\\
		&\quad+C_A\bigg(\!\left(2+\frac{\pi^2}{3}\right)L_h^2-12\zeta_3L_h-12+\frac{\pi^2}{6}+18\zeta_3+\frac{19\pi^4}{90}\bigg)\bigg] + \mathcal{O}(\alpha_s^2) \Bigg\} \,.
\end{align}

The soft function $S_1$ is renormalized multiplicatively. After renormalization, it is simply given by the running $b$-quark mass, such that 
\begin{equation}
\label{eq:O1re}
   S_1(\mu) = Z_{gg}^{-1} Z_{11} S_1^{(0)} 
   = Z_m^{-1} S_1^{(0)} = m_b(\mu) \,.
\end{equation}

\subsection{Form factor in terms of renormalized quantities}
\label{sec:renormFF}

Having all expressions for the renormalized quantities at hand, we can perform the convolution integrals in \eqref{eq:refact} and obtain explicit expressions for the renormalized terms $T_i(\mu)$ (with $i=1,2,3$) up to order $\mathcal{O}(\alpha_s^2)$. We find
\begin{equation}
\begin{aligned}
	\label{eq:T1T2T3}	
	T_1(\mu)&=\mathcal{M}_0\Bigg\{-2+\frac{\alpha_s}{4\pi}\bigg[C_F\bigg(-\frac{\pi^2}{3}L_h^2+(12+8\zeta_3)L_h-36-\frac{2\pi^2}{3}-\frac{11\pi^4}{45}\bigg) \\
	&\quad +C_A\bigg(\left(2+\frac{\pi^2}{3}\right)L_h^2-12\zeta_3L_h-12+\frac{\pi^2}{6}+18\zeta_3+\frac{19\pi^4}{90}\bigg)\bigg] + \mathcal{O}(\alpha_s^2) \Bigg\} \,, \\
	T_2(\mu)&=\mathcal{M}_0\frac{\alpha_s}{4\pi}\bigg[C_F\bigg(\frac{2\pi^2}{3}L_hL_m-\frac{\pi^2}{3}L_m^2+\frac{2\pi^2}{3}+8\zeta_3+\frac{7\pi^4}{45}\bigg)  \\
	&\quad +C_A\bigg(-\frac{2\pi^2}{3}L_hL_m
    +\frac{\pi^2}{3}L_m^2+8\zeta_3L_m-\frac{\pi^2}{2}-6\zeta_3-\frac{\pi^4}{30}\bigg) + \mathcal{O}(\alpha_s^2) \bigg] \,, \\
	T_3(\mu)&=\mathcal{M}_0\Bigg\{\frac{L^2}{2}+\frac{\alpha_s}{4\pi}\bigg[C_F\bigg(-\frac{L^4}{12}-L^3-3L_mL^2+\left(4-\frac{\pi^2}{3}\right)L^2 \\
	&\qquad +\left(\frac{2\pi^2}{3}+8\zeta_3\right)L -8\zeta_3L_m-4\zeta_3-\frac{\pi^4}{9}\bigg) \\
	&\quad +C_A\bigg(-\frac{5L^4}{12}-L_mL^3-\frac{L_m^2L^2}{2}+\left(1+\frac{\pi^2}{12}\right)L^2+4\zeta_3L_m\bigg)\bigg] + \mathcal{O}(\alpha_s^2) \Bigg\} \,.
\end{aligned}
\end{equation}
Adding up the three terms, we reproduce the result for the renormalized form factor given in 
\eqref{eq:ampfrombare}.

\section{RG evolution equations}
\label{sec:RG}

In general, the anomalous dimensions can be extracted from the renormalization factors $Z_{ij}$ defined in \eqref{Zijdef} using the relation
\begin{equation}
	\label{eq:ADdef}
	\gamma_{ij}=2 \alpha_s \frac{\partial}{\partial \alpha_{s}} Z_{i j}^{(1)} \,,
\end{equation}
where $Z_{ij}^{(1)}$ denotes the coefficient of the single $1/\epsilon$ pole in $Z_{ij}$. 

\subsection{Evolution equations for the hard matching coefficients}
\label{subsec:hardRGE}

The renormalized hard functions obey the RG equations 
\begin{equation}
\begin{aligned}
\label{eq:RGEsH2H3}
	\frac{\df}{\df\ln\mu}H_3(\mu)
	&= \gamma_{33}\spac H_3(\mu) \,,\\
	\frac{\df}{\df\ln\mu}H_2(z,\mu)
	&= \int_0^1\df z^\prime\,H_2(z^\prime,\mu)\spac\gamma_{22}(z^\prime,z) \,,\\
	\frac{\df}{\df\ln\mu}\,\braces{\bar{H}_2(z,\mu)}
	&= \int_0^\infty\df z^\prime\,\braces{\bar{H}_2(z^\prime,\mu)}\spac\frac{z}{z^\prime}\spac
	 \braces{\gamma_{22}(z^\prime,z)} \,.
\end{aligned}
\end{equation}
As in the photon case, the evolution equation for $H_1$ is more involved. This is due to various contributions the renormalized hard function $H_1(\mu)$ receives from operator mixing and the ``mismatch contributions'' discussed in section~\ref{subsec:T1ren}. Following the steps laid out in \cite{Liu:2020wbn}, we obtain
\begin{equation}
\label{eq:H1RGE}
   \frac{\df H_{1}(\mu)}{\df \ln \mu} 
   = D_{\mathrm{cut}}(\mu)+\gamma_{11} H_{1}(\mu) + 4\int_0^1\frac{\df z}{z}\!\Big[ \bar{H}_2(z,\mu)\gamma_{21}(z)
    - \braces{\bar{H}_2(z,\mu)}\braces{\gamma_{21}(z)} \Big] \,,
\end{equation}
with
\begin{equation}
	\label{eq:DcutNLO}
	\begin{aligned}
		D_{\text{cut}}(\mu)=-\frac{T_F\spac\alpha_s}{\pi}\frac{y_b(\mu)}{\sqrt{2}}\left[\frac{\alpha_s}{4\pi}\left(C_F-\frac{C_A}{2}\right)16\zeta_3+\mathcal{O}(\alpha_s^2)\right] .
	\end{aligned}
\end{equation}
This quantity exhibits single-logarithmic terms in higher orders, $D_\text{cut}\ni\alpha_s(\alpha_s L_h)^n$ for $n\ge 2$. In order to solve the RG equation for $H_1(\mu)$, it would be necessary to resum these logarithms to all orders. 

\subsection{Evolution equations for the jet and soft functions}
\label{subsec:5.1}

The renormalized jet and soft functions satisfy the RG equations
\begin{equation}
  \label{eq:RGEsOperators}
  \begin{aligned}
  	\frac{\df}{\df\ln\mu}S_1(\mu)&=- (\gamma_{11}-\gamma_{gg})\spac S_1(\mu)\,,\\
  	\frac{\df}{\df\ln\mu}S_2(z,\mu)&=-\int_0^1\df z^\prime\spac 
	 \Big[ \gamma_{22}(z,z^\prime) - \gamma_{gg}\spac\delta(z-z') \Big]\spac
	  S_2(z^\prime,\mu)-\gamma_{21}(z)S_1(\mu)\,,\\
  	\frac{\df}{\df\ln\mu}\braces{S_2(z,\mu)}&=-\int_0^1\df z^\prime\spac
	 \Big[ \braces{\gamma_{22}(z,z^\prime)} - \gamma_{gg}\spac\delta(z-z') \Big]\spac
	 \braces{S_2(z^\prime,\mu)}-\braces{\gamma_{21}(z)}S_1(\mu)\,,\\
  	\frac{\df}{\df\ln\mu}J(p^2,\mu)&=-\int_0^\infty\df x\,\gamma_J(p^2,xp^2)J(xp^2,\mu)\,,\\
  	\frac{\df}{\df\ln\mu}S_3(w,\mu)&=-\int_0^\infty\df w^\prime\,\gamma_S(w,w^\prime)S_3(w^\prime,\mu)\,.
  \end{aligned}
\end{equation}
We collect the relevant expressions for the anomalous dimensions in appendix~\ref{app:ADRG}. Comparing these expressions with the corresponding ones in the photon case, we find two main differences. First, the cusp terms and the convolution kernels of the anomalous dimensions (except for $\gamma_{11}$) do not share the same color factors anymore, leading to highly non-trivial solutions of the corresponding RG equations. Secondly, since the renormalization factors for the soft functions $S_2$ and $S_3$ involve a factor $Z_{gg}^{-1}$ to render them independent of the hard scale $M_h$, the anomalous dimensions for these soft functions receive a contribution from $\gamma_{gg}$ as well. 

From the renormalized form factor \eqref{eq:refact} and the renormalization condition for the soft function $S_3$ we may deduce the non-trivial relation 
\begin{equation}
	\label{eq:ADrelation}
	\big(\gamma_{33}-\gamma_{gg}\big)\,\delta(1-x)=\gamma_{J}\left(\frac{M_{h} w}{\ell_{+}}, x \frac{M_{h} w}{\ell_{+}}\right)+\gamma_{J}\left(-M_{h} \ell_{+},-x M_{h} \ell_{+}\right)+\gamma_{S}(w, w/x) \,,
\end{equation}
which holds to all orders in $\alpha_s$. Despite appearance, the right-hand side of this formula is independent of $\ell_+$ and $w$.

\subsection{Evolution equations for the form factor and its three components}

The renormalized $gg\to h$ form factor fulfills the evolution equation
\begin{equation}
	\label{eq:amprscaledep}
	\frac{\df F_{gg}(\mu)}{\df\ln\mu} = \gamma_{gg} F_{gg}(\mu) \,,
\end{equation}
where 
\begin{equation}
	\gamma_{gg} = \frac{\alpha_s}{4\pi} \left( 4\spac C_A\spac L_h - 2\beta_0 \right) + \mathcal{O}(\alpha_s^2)
\end{equation}
is the anomalous dimension associated with $Z_{gg}$. We may also compute the scale dependence of each of the three terms $T_1(\mu)$, $T_2(\mu)$ and $T_3(\mu)$ individually, finding
\begin{equation}
	\label{eq:scaleT1T2T3}
	\begin{aligned}
		\frac{\df T_1(\mu)}{\df\ln\mu}&=\mathcal{M}_0\Bigg\{\frac{\alpha_s}{4\pi}\left[- \big(C_A-C_F\big)\frac{4\pi^2}{3}L_h+8\zeta_3\big(3C_A-2C_F\big)\right]+\mathcal{O}(\alpha_s^2)\Bigg\}\,,\\
		\frac{\df T_2(\mu)}{\df\ln\mu}&=\mathcal{M}_0\Bigg\{\frac{\alpha_s}{4\pi}\left[\big(C_A-C_F\big)\frac{4\pi^2}{3}L_h-16\zeta_3C_A\right]+\mathcal{O}(\alpha_s^2)\Bigg\}\,,\\
		\frac{\df T_3(\mu)}{\df\ln\mu}&=\mathcal{M}_0\,\Bigg\{\frac{\alpha_s}{4\pi}\left[2L^2\left(C_A L_h-\frac{\beta_0}{2}\right)+16\zeta_3\left(C_F-\frac{C_A}{2}\right)\right]+\mathcal{O}(\alpha_s^2)\Bigg\}\,.
	\end{aligned}
\end{equation}

\section[Large logarithms in the three-loop $gg\to h$ amplitude]{\boldmath Large logarithms in the three-loop $gg\to h$ amplitude}
\label{sec:LL}

Given the RG equations and anomalous dimensions for the ingredients in the factorization formula, we are able to predict the four leading logarithms in the three-loop expression for the $gg\to h$ form factor in analytic form. To this end, we solve the evolution equations iteratively and determine the leading large logarithms in the hard matching coefficients and the soft functions at NNLO in perturbation theory. This is discussed in detail in appendix~\ref{app:higherlogterms}. As in the photon case studied in \cite{Liu:2020ydl}, we convert our results to the on-shell scheme. Therefore, we first express the running parameters $m_b(\mu)$ and $y_b(\mu)$ in terms of the pole mass $m_b$. We then eliminate the remaining scale dependence by taking $\mu^2=\hat\mu_h^2\equiv-M_h^2-i0$. This greatly simplifies the three-loop expressions. At NNLO, we find (here $v$ denotes the vacuum expectation value of the Higgs field)
\begin{align}
	\label{eq:AmpN3LO}
		F_{gg}(\hat\mu_h) &= T_F\spac\delta_{ab}\frac{\alpha_s(\hat\mu_h)}{\pi}\frac{m_b^2}{v}\,\Bigg\{-2+\frac{L^2}{2}+\frac{\alpha_s(\hat\mu_h)}{4\pi}\Bigg[\frac{C_A-C_F}{12}L^4-C_FL^3 \notag\\
		&\quad+\left(\!\left(1+\frac{5\pi^2}{12}\right)C_A-\frac{2\pi^2}{3}C_F\right)L^2+\left(\!\left(12+\frac{2 \pi ^2}{3}+16 \zeta_3\right)C_F-12\zeta_3\spac C_A\right)L \notag\\
		&\quad+\left(4 \zeta_3-\frac{\pi ^4}{5}-20\right)C_F+\left(12 \zeta_3+\frac{8 \pi ^4}{45}-\frac{\pi ^2}{3}-12\right)C_A\Bigg] \notag\\
		&\quad +\left(\frac{\alpha_s(\hat\mu_h)}{4\pi}\right)^2 \Bigg[\frac{(C_A-C_F)^2}{90}L^6+(C_A-C_F)\left(\frac{\beta_0}{30}-\frac{C_F}{10}\right)L^5 \notag\\
		&\quad +d_{4}^{\text{OS}}L^4+d_{3}^{\text{OS}}L^3+\cdots \Bigg]\Bigg\}\,,
\end{align}
where $L=\ln[(-M_h^2-i0)/m_b^2]$, and 
\begin{equation}
	\label{eq:dOS}
	\begin{aligned}
		d_{4}^{\text{OS}}&= \left(\frac{3}{2}+\frac{\pi ^2}{18}\right)C_F^2-\left(\frac{191}{54}+\frac{\pi ^2}{24}\right)C_FC_A+\left(\frac{85}{108}-\frac{\pi ^2}{72}\right)C_A^2+\frac{32 C_F-5 C_A}{27}T_Fn_f \,,\\
		d_{3}^{\text{OS}}&= \left(\frac{20 \zeta_3}{3}+\frac{7 \pi ^2}{9}-\frac{1}{2}\right)C_F^2-\left(10 \zeta_3+\frac{235}{18}+\frac{43 \pi ^2}{27}\right)C_FC_A\\
		&\quad +\bigg(\frac{10 \zeta_3}{3}+\frac{11 \pi ^2}{18}+\frac{4}{3}\bigg)C_A^2+\left(\frac{22}{9}+\frac{8 \pi ^2}{27}\right) C_F T_Fn_f-\left(\frac{2}{3}+\frac{2 \pi ^2}{9}\right) C_A T_Fn_f\,.
	\end{aligned}
\end{equation}
The coefficients of the color structures $C_F^2$ and $C_F T_F$ agree with the corresponding coefficients in the photon case.

\section{Resummation}
\label{sec:resum}

In this section, we want to resum the large logarithms to all orders in perturbation theory. We need therefore solve the RG equations for the different hard, jet, and soft functions. Choosing to set the scale where we evaluate our predictions as $\mu=\mu_h$, all large logarithms in the evolution of the hard functions vanish, leaving them in the evolution of the jet and soft functions. In this context, the general logarithmic structure reads:
\begin{equation}
    \label{eq:logstructure}
    \begin{aligned}
        T_1(\mu_h) &= T_F\spac\delta_{ab}\frac{y_b(\mu_h)}{\sqrt{2}}\frac{\alpha_s(\mu_h)}{\pi}m_b(\mu_h)\left[-2+\sum_{n\geq 1}\alpha_s(\mu_h)^n\,a_n\right]\,,\\
        T_2(\mu_h) &= T_F\spac\delta_{ab}\frac{y_b(\mu_h)}{\sqrt{2}}\frac{\alpha_s(\mu_h)}{\pi}m_b(\mu_h)\sum_{n\geq 1}\alpha_s(\mu_h)^n\sum^{n+1}_{i=0}b_{n,i}\,L^i\,,\\
        T_3(\mu_h) &= T_F\spac\delta_{ab}\frac{y_b(\mu_h)}{\sqrt{2}}\frac{\alpha_s(\mu_h)}{\pi}m_b(\mu_h)\sum_{n\geq 0}\alpha_s(\mu_h)^n\sum^{2n+2}_{i=0}c_{n,i}\,L^i,
    \end{aligned}
\end{equation}
where $a_n\,,c_{n,i}$ and $c_{n,i}$ are numbers. It is obvious that $T_3$ dominates the logarithmic corrections since it is of Sudakov type. Hence in the following, we will only focus on the third term. The photon case has been resummed to next-to-leading double-logarithmic accuracy (NLL) in \cite{Liu:2020tzd,Liu:2020wbn}. In this paper, we include one more tower of logarithms, i.e.\ we resum factors of $\alpha_s^n L^{2n}$, $\alpha_s^n L^{2n-1}$ and $\alpha_s^n L^{2n-2}$ to all orders of perturbation theory. This is conventionally named NLL$'$ accuracy. 

In the literature, one distinguishes two different schemes for the resummation of large logarithms in Sudakov problems. The so-called ``RG-improved perturbation theory'' rests on the assumption that $\alpha_s\spac L=\mathcal{O}(1)$, where $L$ is the large logarithm in a given problem. The parametrically leading terms in the {\em logarithm\/} of a quantity are then of order $L(\alpha_s\spac L)^n\sim \alpha_s^{-1}(\alpha_s\spac L)^n$ and are formally larger than $\mathcal{O}(1)$. The leading-order approximation (LO) is therefore defined by the simultaneous resummation of all terms of order $L(\alpha_s\spac L)^n$ and $(\alpha_s\spac L)^n$ in the logarithm of the quantity; i.e., all such logarithms get exponentiated in the expression for the quantity itself. The NLO approximation resums in addition the terms of order $\alpha_s(\alpha_s\spac L)^n$ in the exponent, and so on. In the double-logarithmic counting scheme, instead, one assumes that $\alpha_s\spac L^2=\mathcal{O}(1)$. In this case the resummation is performed for the observable itself. In the leading double-logarithmic approximation (LL), all terms of order $\alpha_s^n\spac L^{2n}$ are resummed. At the next order (NLL), one resums the logarithms of the form $\alpha_s^n\spac L^{2n-k}$ with $k=0,1$, and so on. In table~\ref{tab:T3log} we summarize the ingredients needed at a given order in the two schemes. N$^{k+1}$LL resummations (with $k\geq 0$) are contained in RG-improved perturbation theory at N$^k$LO, while N$^{k+1}$LL$'$ resummation includes matching corrections at one order higher, however, the same-order anomalous dimensions are used. Hence it is enough to use RG-improved LO jet and soft functions to account for NLL$'$ corrections from the anomalous dimensions. On top of that, it turns out that only constant terms at NLO in the hard, jet, and soft functions at their respective matching scales contribute to the large logarithms at NLL$'$, which simplifies the calculation a lot. 

\begin{table}[hbt]
  \centering
  \begin{tabular}{|c|c|c|c|c|c|}
    \hline
    \hline
    RG-impr. PT & Log. approx. & $\Gamma_{\text{cusp}}\,,\beta$ & $\gamma$ & $H_3\,,S_3\,,J$ & $\alpha_s^n\,L^k$\\
    \hline
    $-$ & LL & LO & $-$ & LO & $k=2n$ \\
    \hline 
    LO & NLL & NLO & LO & LO & $2n-1\leq k\leq 2n$ \\
    \hline 
    $-$ & NLL$'$ & NLO & LO & NLO & $2n-2\leq k\leq 2n$ \\
    \hline 
    NLO & NNLL & NNLO & NLO & NLO & $2n-3\leq k\leq 2n$\\
    \hline
    \hline
  \end{tabular}
  \caption{\label{tab:T3log}  Naming schemes for logarithmic accuracy in $T_3(\mu)$. We list perturbative orders of the cusp anomalous dimension, non-cusp anomalous dimensions $\gamma$, QCD $\beta$ function, and matching functions to obtain resummation at a given logarithmic order.  }
\end{table}

The solution to the RG equation for the jet function has been presented in \cite{Liu:2021mac} to RG-improved LO. In the following, we will first derive the RG-improved soft function at LO. Subsequently, we resum the first three towers of large logarithms in the third term of the amplitude. Note that at NLL$^\prime$ accuracy, there are no contributions from the first and second term apart from the fixed $n=1$ contribution in the second term, which therefore does not need to be resummed at the given logarithmic order. We leave the resummation of further subleading logarithms in the first and second term for future work.

\subsection[RG-improved LO soft function \texorpdfstring{$S_3$}{Lg}]{RG-improved LO soft function \texorpdfstring{\boldmath$S_3$}{Lg}}
\label{subsec:RGsoft}

The RG-improved LO soft function $S_3$ can be derived in a similar manner as has been the soft function of $h\to\gamma\gamma$ in \cite{Liu:2020eqe}. There a general ansatz has been presented via transformation to Laplace space. For our factorization theorem, we may apply the same techniques, which is why we do not recapitulate the whole derivation here again. A major difference is, however, that in our non-abelian scenario for the anomalous dimension of the soft function $\gamma_S$ the cusp term and the non-local convolution kernel do not share the same color factor. This has also been observed for the jet function in \cite{Liu:2021mac} and prevented the calculation of the RG-improved jet function beyond the leading order. Defining the ratio $r_\Gamma=(C_F-C_A/2)/(C_F-C_A)$, we find for the soft function at leading order
\begin{equation}
	\label{eq:SRGiLO}
	\begin{aligned}
		S_3^{\text{LO}}(w,\mu)
        &= U_S(w;\mu_s,\mu)\int_0^\infty\frac{\df w^\prime}{w^\prime}S^{\text{LO}}(w^\prime,\mu_s) \\
        &\hspace{3.2cm} \times I^{1,1}_{2,2}\left(
         \begin{matrix} (-a_{\Delta_\Gamma},1,2r_\Gamma) & ~,~ & (1-a_{\Delta\Gamma},1,2r_\Gamma) \\
                             (1,1,2r_\Gamma) & ~,~ & (0,1,2r_\Gamma)
         \end{matrix}\bigg|\frac{w^\prime}{w}\right) ,
	\end{aligned}
\end{equation}
with
\begin{equation}
	\label{eq:US}
	\begin{aligned}
		U_S(w;\mu_s,\mu)&=\left(\frac{we^{-4r_\Gamma\gamma_E}}{\mu_s^2}\right)^{-a^{(0)}_{\Delta\Gamma}(\mu_s,\mu)}\exp\Big[2S^{(0)}_{\Delta\Gamma}(\mu_s,\mu)+a^{(0)}_{\gamma_s}(\mu_s,\mu)\Big]\,,\\
		 S^{\text{LO}}(w,\mu_s)&=-T_F\spac\delta_{ab}\frac{\alpha_s(\mu_s)}{\pi} m_b(\mu_s) \theta(w-m_b^2).
	\end{aligned}
\end{equation}
Here, $S_3(w,\mu_s)$ denotes the soft function at the matching scale $\mu_s$; $S_{\Delta\Gamma}^{(0)}$, $a_{\Delta\Gamma}^{(0)}$ and $a_{\gamma_s}^{(0)}$ are leading terms of the corresponding RG functions. Their definition and behavior are studied in more detail in appendix \ref{subapp:RGfuncs}. For the sake of intelligibility, we have suppressed the arguments of these functions. The function $I^{1,1}_{2,2}\left(\cdots|x\right)$ is a so-called \textit{Rathie-I} function, defined as
\begin{equation}
    \label{eq:Rathiedef}
    \begin{aligned}
    I^{m,n}_{p,q}\left(\begin{matrix}(a_1,\alpha_1,A_1),\dots,(a_p,\alpha_p,A_p)\\(b_1,\beta_1,B_1),\dots,(b_q,\beta_q,B_q)\end{matrix}\bigg|z\right)=\frac{1}{2\pi i}\int_L\phi(s) z^s\df s\,,\\
    \text{with}\quad \phi(s)=\frac{\prod\limits_{j=1}^m \Gamma^{B_j}(b_j-\beta_js)\prod\limits_{j=1}^n\Gamma^{A_j}(1-a_j+\alpha_j s)}{\prod\limits_{j=m+1}^q \Gamma^{B_j}(1-b_j+\beta_js)\prod\limits_{j=n+1}^p\Gamma^{A_j}(a_j-\alpha_j s)}\,.
    \end{aligned}
\end{equation}
Its definition and properties were first presented in \cite{Rathie}. It is a generalization of the \textit{Meijer-G} function $G^{m,n}_{p,q}$ and related via
\begin{equation}
    \label{eq:MeijerRathie}
    G^{m,n}_{p,q}\left(\begin{matrix}a_1,\dots,a_p\\b_1,\dots,b_q\end{matrix}\bigg|z\right)=I^{m,n}_{p,q}\left(\begin{matrix}(a_1,1,1),\dots,(a_p,1,1)\\(b_1,1,1),\dots,(b_q,1,1)\end{matrix}\bigg|z\right)\,.
\end{equation}

Though the analytic solution takes a rather complicated form, the asymptotic behavior is fairly simple:
\begin{equation}
	\label{eq:asymsoft}
	\begin{aligned}
		S^{\text{LO}}(w,\mu) = S^{\text{LO}}(w,\mu_s) U_S(w;\mu_s,\mu)\left(\frac{\Gamma(1+a^{(0)}_{\Delta\Gamma}(\mu_s,\mu))}{\Gamma(1-a^{(0)}_{\Delta\Gamma}(\mu_s,\mu))}\right)^{2r_\Gamma}+\mathcal{O}(m_b^2/w).
	\end{aligned}
\end{equation}
We have found that only the region above the hyperbola $\ell_-\ell_+>m_b^2$ contributes to the NLL$'$ accuracy. In this context, further corrections from the \textit{Rathie-I} function are not relevant for NLL$'$ resummation, but will come into play in RG-improved perturbation theory. This is however beyond the scope of this paper.

\subsection{Large logarithms at NLL\texorpdfstring{$^\prime$}{Lg} in the form factor}

The scale dependence of the $gg\to h$ form factor is governed by the evolution equation~\eqref{eq:amprscaledep}. It is not scale-invariant due to the external gluon states. At LO in RG-improved perturbation theory, we find \cite{Liu:2020tzd}
\begin{equation}
	\label{eq:AmpRGiLO}
F_{gg}^R(\mu)=e^{2S_{\Gamma_A}(\mu_h,\mu)}\frac{\alpha_s(\mu)}{\alpha_s(\mu_h)}F_{gg}^R(\mu_h)\,,
\end{equation}  
and $\Gamma_A$ stands for the cusp anomalous dimension in the adjoint representation. The scale $\mu_h^2=-M_h^2-i0$ is chosen such that there are no large logarithms left in the hard matching coefficients. The derivation of $F_{gg}^R(\mu_h)$ is highly non-trivial and will be carried out in multiple steps. There are two kinds of contributions. One stems from the RG evolution of the component functions, which is controlled by the respective anomalous dimension. The second one is NLO corrections in these functions at their matching scales. 

The contribution from RG evolution is given by taking the RG-improved LO component function for $T_3$,
\begin{equation}
\label{eq:T3RGiLO}
\begin{aligned}
   T_3^{\text{LO}}(\mu_h) 
   &= \lim_{\sigma\to -1} H_3(\mu_h)^\text{LO} \int_0^{M_h}\!\frac{\df\ell_-}{\ell_-}
    \int_0^{\sigma M_h}\!\frac{\df\ell_+}{\ell_+} \\
   &\quad\times J^{\text{LO}}(-M_h\ell_-,\mu_h)\spac J^{\text{LO}}( M_h\ell_+,\mu_h)\spac S_3^{\text{LO}}(\ell_-\ell_+,\mu_h)\bigg|_{\text{leading\,power}}\,.
\end{aligned}
\end{equation} 
In principle, the matching scales of the two jet functions could be different, since they depend on different dynamical scales $\ell_\pm$. They are chosen such that all logarithms are located only in the evolution factors. The LO soft function has been derived in the previous section, the jet function is given by
\begin{equation}
\label{eq:JetRGiLO}
\begin{aligned}
   J^{\text{LO}}(p^2,\mu)
   &= \left(\frac{-p^2}{\mu_j^2}\right)^{a^{(0)}_{\Delta\Gamma}(\mu_j,\mu)}
    \exp\big[-2S^{(0)}_{\Delta\Gamma}(\mu_j,\mu)-2r_\Gamma\,\gamma_E\,a^{(0)}_{\Delta\Gamma}(\mu_j,\mu)\big] \\
   &\quad\times \left(\frac{\Gamma(1-a^{(0)}_{\Delta\Gamma}(\mu_j,\mu))}{\Gamma(1+a^{(0)}_{\Delta\Gamma}(\mu_j,\mu))}
    \right)^{r_\Gamma} \,,
\end{aligned}
\end{equation}
and was first presented in \cite{Liu:2021mac}. Here $\Delta\Gamma$ stands for the difference between the cusp anomalous dimension in the fundamental and adjoint representation. 

To extract the first three towers of large logarithms, we only need to enter the regime $\ell_+\ell_-\gg m_b^2$. We may therefore use the asymptotic expression for the soft function $S_3$ given by \eqref{eq:asymsoft}. In the first step, we define the following abbreviations
\begin{equation}
	\label{eq:as}
	a_s=a_{\Delta\Gamma}^{(0)}(\mu_s,\mu_h),\,\,\,a_-=a_{\Delta\Gamma}^{(0)}(\mu_-,\mu_h),\,\,\,a_+=a_{\Delta\Gamma}^{(0)}(\mu_+,\mu_h)\,,
\end{equation}
where $\mu_-$ is the matching scale entering the jet function $J(-M_h\ell_-,\mu_h)$ while $\mu_+$ is that entering the jet function $J(M_h\ell_+,\mu_h)$. The factors of gamma functions in the RG-improved jet \eqref{eq:JetRGiLO} and soft \eqref{eq:SRGiLO} functions can be further expanded to 
\begin{equation}
	\left[e^{4\gamma_E a_s}\,\frac{\Gamma^2(1+a_s)}{\Gamma^2(1-a_s)}e^{2\gamma_E a_-}\,\frac{\Gamma(1+a_-)}{\Gamma(1-a_-)}e^{2\gamma_E a_+}\,\frac{\Gamma(1+a_+)}{\Gamma(1-a_+)}\right]^{r_\Gamma}=1+\mathcal{O}(a_s^3,a_-^3,a_+^3)\,.
\end{equation}
The jet and soft functions must be free of large logarithms at the matching scales $\mu_\pm$ and $\mu_s$. Since these functions are integrated over soft ($\ell_+\ell_-\sim m_b^2$) and hard ($\ell_+\ell_-\sim M_h^2$) regions, we must set these matching scales dynamically under the integral. Hence we fix $\mu_s^2=\ell_-\ell_+$, $\mu_-^2=\sigma M_h\ell_-$ and $\mu_+^2= M_h\ell_+$. Additionally, the prefactor $\alpha_s(\mu_s)$ entering the soft function (see \eqref{eq:US}) should be converted into a scheme that only depends on the hard scale
\begin{equation}
	\label{eq:alphas}
	\begin{aligned}
		\alpha_s(\nu)=\frac{\alpha_{s}\left(\mu\right)}{X}\left[1-\frac{\alpha_{s}\left(\mu\right)}{4 \pi} \frac{\beta_{1}}{\beta_{0}} \frac{\ln X}{X}+\mathcal{O}(\alpha_s^2)\right],~\text{with}~ X=1-\frac{\alpha_s(\mu)}{4\pi}\beta_0\ln\frac{\mu^2}{\nu^2},
	\end{aligned}
\end{equation} 
and we abbreviate the logarithms as follows when necessary
\begin{equation}
	\label{eq:Ls}
	L_-=\ln\frac{\mu_h^2}{\mu_-^2}\,,\,L_+=\ln\frac{\mu_h^2}{\mu_+^2}\,,\,L_s=\ln\frac{\mu_h^2}{\mu_s^2}=L_-+L_+\,,\,\text{and}\,\, L=\ln\frac{\mu_h^2}{m_b^2}\,.
\end{equation} 
The relevant parameter $\rho$ in NLL$'$ resummation is defined as
\begin{equation}
    \label{eq:rhodef}
    \r=\frac{\a(\mu_h)}{4\pi}\frac{\dg}{2}L^2=\frac{\a(\mu_h)}{2\pi}(C_F-C_A)L^2\sim -1.192 + 0.955\,i\,,
\end{equation}
Substituting $L_+=xL$, $L_-=yL$, we find up to order NLL$^\prime$
\begin{equation}
\label{eq:alphaconcersionFull}
	\a(\mu_s)=\a(\mu_h)\left(1+\frac{\r}{L}\frac{2\pb}{\dg}(x+y)+\frac{\r^2}{L^2}\frac{4\pbs}{(\dg)^2}(x+y)^2+\mathcal{O}(L^{-3})\right)\,.
\end{equation}
Here, $\pb=\b$ and the coloring is related to a comparison with the resummation of the photon case and will be explained further later on. 

As mentioned before, there are also contributions from the NLO corrections at the matching scales. Due to the dynamic scale setting, logarithms at the matching scales vanish. Hence the corrections from the hard and jet functions are given by the constant terms of these functions. For the soft function though, in principle there are some extra functional terms, see \eqref{eq:softrenormalizedcont}. However, all these terms go to zero when $\hat{w}$ is large, such that their contributions are not relevant here. We find for the combined contribution at the matching scales
\begin{equation}
    \label{eq:matchingcorr}
    \begin{aligned}
        \Delta_\text{matching} =  \frac{\rho}{L^2}\frac{2}{\Delta\Gamma_0}\left[C_F\left(8-\frac{2\pi^2}{3}\right)+C_A\left(2+\frac{\pi^2}{6}\right)\right].
    \end{aligned}
\end{equation}
Adding all contributions together, $T_3(\mu_h)$ reads
\begin{equation}
\label{eq:T3LO}
\begin{aligned}
   & T_3(\mu_h)|_{\text{NLL}'} = \mathcal{M}_0(\mu_h)L^2\int_0^1\!\df x \int_0^{1-x}\!\df y
    \left[1+\frac{\r}{L}\frac{2\pb}{\dg}(x+y)+\frac{\r^2}{L^2}\frac{4\pbs}{(\dg)^2}(x+y)^2\right] \\
   &\quad\times \left\{1+\frac{\rho}{L^2}\frac{2}{\Delta\Gamma_0}\left[C_F\left(8-\frac{2\pi^2}{3}\right)
    + C_A\left(2+\frac{\pi^2}{6}\right)\right]\right\} \\
   &\quad\times \exp\Big[2S_{\Delta\Gamma}^{(0)}(\mu_s,\mu_h) - 2S_{\Delta\Gamma}^{(0)}(\mu_-,\mu_h)
    - 2S_{\Delta\Gamma}^{(0)}(\mu_+,\mu_h)+a^{\text{(0)}}_{\gamma_s}(\mu_s,\mu_h)
    + a^{\text{(0)}}_{\gamma_m}(\mu_s,\mu_h)\Big]_{\text{NLL}'} \,,
\end{aligned}
\end{equation}
where the term in square brackets accounts for the contribution from converting the strong coupling constant in the prefactor, the term in curly braces is generated by corrections to the component functions at the matching scale, and the exponential factor is due to scale evolution. We insert the expressions for the RG functions from \ref{app:ADRG} and perform all remaining integrals. Neglecting terms of order $\mathcal{O}(L^{-3})$ we arrive at
\begin{equation}
\label{eq:T3atNLLprime}
	\begin{aligned}
		T_3(\mu_h)|_{\text{NLL}^\prime} =& \mathcal{M}_0(\mu_h)\frac{L^2}{2}\sum_{n=0}^\infty(-\r)^n\frac{2\Gamma(n+1)}{\Gamma(2n+3)}\bigg\{1+\frac{1}{L}\bigg[\r\frac{-(\g_s^0+\g_m^0)+2\pb}{\dg}\frac{2n+2}{2n+3}\\
  &-\r^2\frac{\b}{\dg}\frac{(n+1)^2}{(2n+3)(2n+5)}\bigg]+\frac{1}{L^2}\bigg[\r\frac{C_F\left(4-\frac{\pi^2}{3}\right)+C_A\left(1+\frac{\pi^2}{12}\right)}{C_F-C_A}\\
		&+\r^2\bigg(-\frac{\b(\g_s^0+\g_m^0)}{(\dg)^2}\frac{n+1}{n+2}-\frac{\Dg}{(\dg)^2}\frac{(n+1)^2}{(n+2)(2n+3)}\\
		&+\frac{(\g_s^0+\g_m^0)^2}{(\dg)^2}\frac{n+1}{2(n+2)}-\frac{\pb(\g_s^0+\g_m^0)}{(\dg)^2}\frac{2(n+1)}{n+2}+\frac{\pbs}{(\dg)^2}\frac{4(n+1)}{n+2}\bigg)\\
		&+\r^3\bigg(\frac{\b(\g_s^0+\g_m^0)}{(\dg)^2}\frac{(n+1)^2}{2(n+3)(2n+3)}-\frac{\b^2}{(\dg)^2}\frac{(n+1)^2(7n+18)}{6(n+3)(2n+3)(2n+5)}\\
		&-\frac{\pbs}{(\dg)^2}\frac{(n+1)^2}{(n+3)(2n+3)}\bigg)+\r^4\frac{\b^2}{(\dg)^2}\frac{(n+1)^2(n+2)}{8(n+4)(2n+3)(2n+5)}\bigg]\bigg\}\,.
	\end{aligned}
\end{equation}
Note that $\g_s^0=-6C_F+2\pb$. The first two towers of logarithms (up to order $\mathcal{O}(L^{-1})$) have already been derived in \cite{Liu:2020tzd,Wang:2021vtp}. As a non-trivial cross-check expression \eqref{eq:T3atNLLprime} reproduces correctly the leading logarithms in the three-loop amplitude \eqref{eq:AmpN3LO}. In \cite{ResummationPaper}, the resummed amplitude for the $h\to\g\g$ process was presented at NLL$^\prime$ accuracy. To compare this with our result \eqref{eq:T3atNLLprime}, it is not sufficient to set $C_A\to0$. The reason for that is that the prefactor of our $gg\to h$ process features a strong coupling constant evaluated at the soft scale which is subject to being converted to an evaluation at the high scale \eqref{eq:alphas} and therefore gives rise to additional terms suppressed by one and two factors of $1/L$, see \eqref{eq:alphaconcersionFull}. In contrast, in the $h\to\g\g$ case the prefactor is $\alpha_b(\mu_s)=(Q_b e)/(4\pi)$, which is related to the QED coupling constant at the high scale via $\alpha_b(\mu_s)=\alpha_b(\mu_h)(1+\mathcal{O}\left(\alpha_b(\mu_h)\right)$. To account for this effect, we must consequently set $C_A\to0$ and $\pb\to0$ while keeping $\b\neq0$. Hence, we colored the corresponding $\pb$-terms to easily allow comparison between abelian and non-abelian processes. Note that up to NLL, the $gg\to h$ amplitude can be retrieved from the $h\to\g\g$ amplitude by a simple exchange of color factors $C_F\to C_F-C_A$.

The series in \eqref{eq:T3atNLLprime} can be cast into more elegant form by executing the infinite sums. We introduce the special functions
\begin{equation}
    \label{eq:specialfunc}
    \begin{aligned}
        F_1(z) &= {}_2F_2\left(1,1;\frac{3}{2},2;-\frac{z}{4}\right)\,,\\
        F_2(z) &= {}_2F_2\left(1,1;\frac{1}{2},2;-\frac{z}{4}\right)\,,\\
        D(z) &= e^{-z^2}\int_0^z \df x\,e^{\,x^2}\,,
    \end{aligned}
\end{equation}
where $D(z)$ is a so-called Dawson function. We obtain 
\begin{equation}\label{eq:T3specialsimple}
\begin{aligned}
   T_3(\mu_h)\big|_{\text{NLL}^\prime}
   &= \mathcal{M}_0(\mu_h)\frac{L^2}{2} 
    \bigg\{ F_1(\r)+\frac{1}{L}\frac{2}{\dg}\bigg[4\pb-3\beta_0-2\left(\gamma_s^0+\gamma_m^0\right) \\
   &\qquad + \big(-2(4\pb-3\beta_0)+\r\beta_0+4\left(\gamma_s^0+\gamma_m^0\right)\big)\,
    \frac{D\left(\frac{\sqrt{\r}}{2}\right)}{\sqrt{\r}}\bigg] \\
   &\quad +\frac{1}{L^2}\frac{1}{(\dg)^2}\,\bigg[ \bigg(-\frac{\r^2}{4}\beta_0^2+\frac{\r}{6}\left(24\pbs-7\beta_0^2\right)-2\r\beta_0\left(\gamma_s^0+\gamma_m^0\right) \\
   &\qquad +18\beta_0^2+4\Dg\bigg)\sqrt{\r}D\left(\frac{\sqrt{\r}}{2}\right)\\
   &\qquad + \left((4+\r)\beta_0^2-8\left(\gamma_s^0+\gamma_m^0\right)(2\pb-\beta_0)+4\left(\gamma_s^0+\gamma_m^0\right)^2\right)\frac{\r}{4}\\
   &\qquad - \left(6\beta_0^2-2\left(\gamma_s^0+\gamma_m^0\right)(2\pb-\beta_0)+\left(\gamma_s^0+\gamma_m^0\right)^2\right)\r F_2(\r)\\
   &\qquad - \bigg[ 4\beta_0^2+2\Dg
    + \frac{C_A\left(\frac{\pi^2}{12}+1\right)-C_F\left(\frac{\pi^2}{3}-4\right)}{C_A-C_F}\,(\dg)^2 \bigg]\,\r F_1(\r)\bigg] \bigg\} \,.
\end{aligned}
\end{equation}

For a better intelligibility of the resummed result \eqref{eq:T3specialsimple}, we find it instructive to give the asymptotic behavior of the special functions. In the limits $\rho\to0,\infty$, the hypergeometric functions can be expanded as
\begin{align}
	\begin{aligned}
		F_1(\rho) &= \left\{ 
		\begin{aligned}
			& 1-\frac{\rho }{12}+\frac{\rho ^2}{180}-\frac{\rho ^3}{3360}+O\left(\rho ^4\right)\,,&\rho\to 0\,,\\
			& 2\frac{\ln \left(\rho e^{\gamma_E}\right)}{\rho}-\frac{4}{\rho^2}+\mathcal{O}(\rho^{-3})\,,&\rho\to \infty\,,
		\end{aligned} \right.
	\end{aligned}\\
	\begin{aligned}
		F_2(\rho) &= \left\{ 
		\begin{aligned}
			& 1-\frac{\rho }{4}+\frac{\rho ^2}{36}-\frac{\rho ^3}{480}+O\left(\rho ^4\right)\,,&\rho\to 0\,,\\
			& \frac{4-2\ln \left(\rho e^{\gamma_E}\right)}{\rho}+\frac{12}{\rho^2}+\mathcal{O}(\rho^{-3})\,,&\rho\to \infty\,.
		\end{aligned} \right.
	\end{aligned}
\end{align}
The Dawson function appearing first at NLL obeys the following behavior
\begin{equation}
	\label{eq:D}
	D\left(\frac{\sqrt{\rho}}{2}\right) = \left\{ 
	\begin{aligned}
		& \frac{\sqrt{\rho}}{2}\left[1-\frac{\rho }{6}+\frac{\rho ^2}{60}-\frac{\rho^3}{840}+O\left(\rho ^4\right)\right]\,,&\rho\to 0\,,\\
		& \frac{1}{\sqrt{\rho}}\left[1+\frac{2}{\rho }+\frac{12}{\rho ^2}+\mathcal{O}(\rho^{-3})\right]\,,&\rho\to \infty\,.
	\end{aligned} \right.
\end{equation}

\begin{figure}[t]
	\begin{center}
		\includegraphics[width=0.45\textwidth]{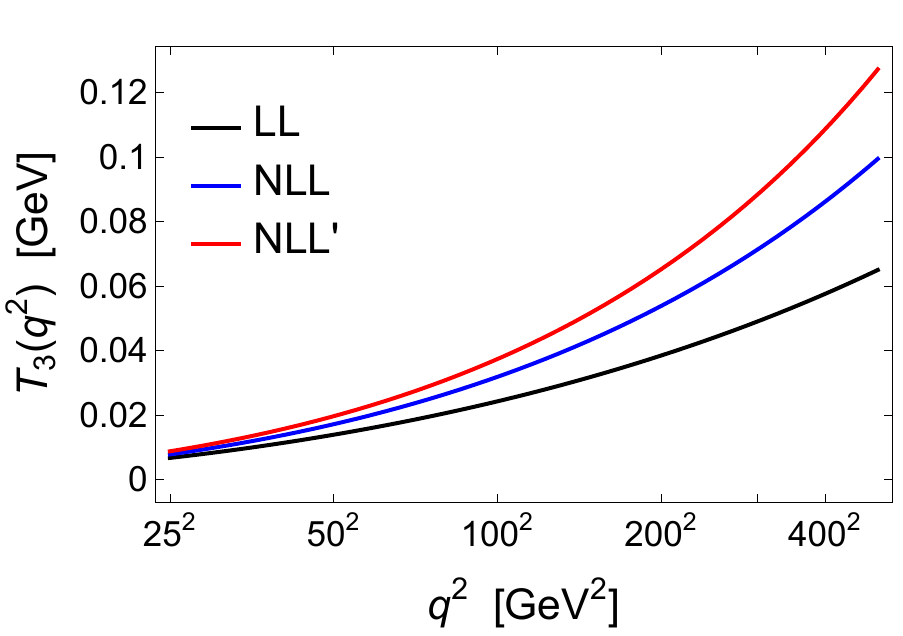} \\
		\includegraphics[width=0.45\textwidth]{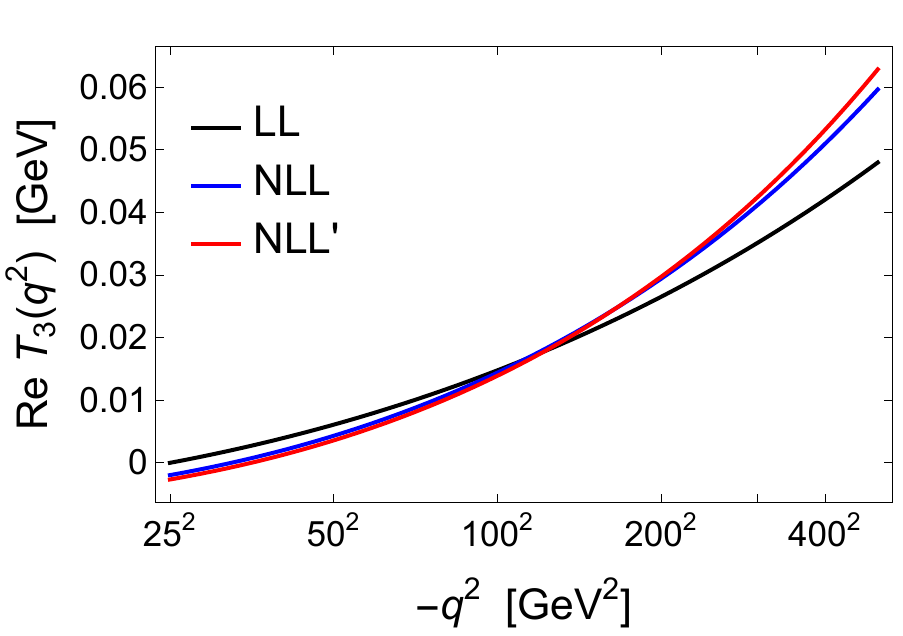} 
		\includegraphics[width=0.45\textwidth]{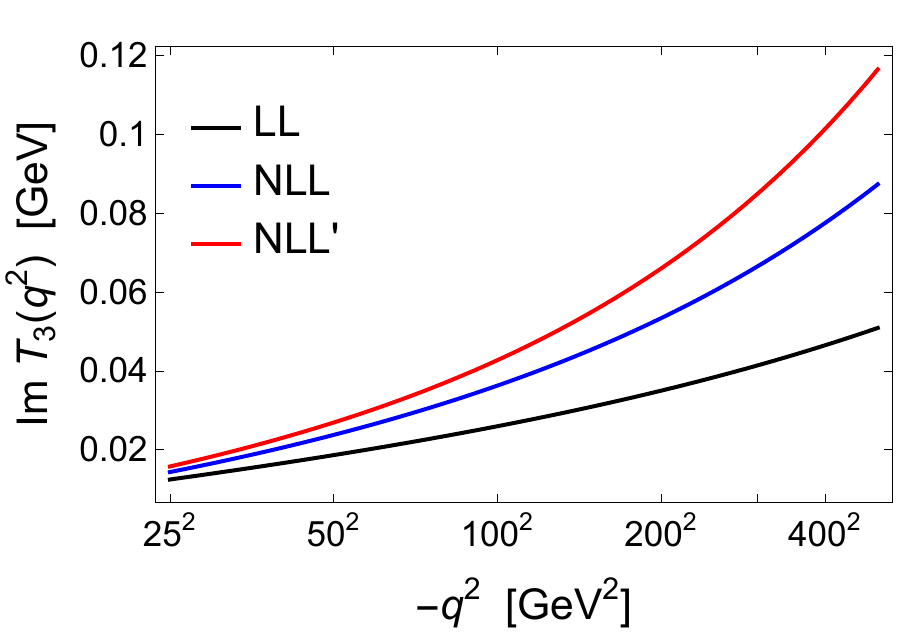}
		\caption{\label{fig:T3musq} Resummed $T_3$ at LL (black), NLL (blue) and NLL$^\prime$ (red) accuracy. We fix the strong coupling constant at $\alpha_s(M_h)$ and vary the hard scale $\mu_h^2=q^2$ entering the large logarithms $L$ and expansion parameter $\rho$. The upper panel shows $T_3$ for $q^2>0$, the lower two panels give the real and imaginary part for $q^2<0$. NLL$(^\prime)$ corrections become increasingly more important for $q^2$-values further away from its physical value $q^2=-M_h^2$.}
	\end{center}
\end{figure}

In figure \ref{fig:T3musq} we show the resummed $T_3$ at LL (black), NLL (blue) and NLL$^\prime$ (red) accuracy. Here, we fix the strong coupling constant at $\alpha_s(M_h)$ and vary the hard scale $\mu_h^2\equiv q^2$ entering the large logarithms $L$ and expansion parameter $\rho$. We give the plots for both $q^2>0$ (upper panel) and real and imaginary part for $q^2<0$ (lower panels). NLL$(^\prime)$ corrections become increasingly more significant the further one takes $q^2$ from its physical value $q^2=-M_h^2$ chosen in the resummation.

\section{Conclusions}
\label{sec:con}

In this work, we have successfully used SCET to derive the factorization theorem for the Higgs-boson production process $gg\to h$ via light quark loops. We followed the steps of \cite{Liu:2019oav, Liu:2020wbn}, where the methodology was applied to the Higgs decay $h\to\gamma\gamma$ via a light quark loop. This has been achieved at the bare level by adopting the RBS scheme. In this way, we are able to write the bare factorization theorem such that no endpoint divergences occur, without the need to introduce an additional regulator apart from dimensional regularization. This is possible by the use of two refactorization conditions that relate component functions of the second term of the factorization theorem that are in the endpoint region to those of the third term. This procedure subtracts the divergent parts in between the two terms. However, since the ``infinity-bin" contribution is subtracted twice, it must be added back as a further contribution to the first hard matching coefficient. We highlight that in contrast to the abelian photon case, an additional scale is involved, namely the QCD confinement scale $\Lambda_{QCD}$ where non-perturbative effects come into play. We must therefore match the amplitude at hand to the gluon operator $\langle O_{gg}\rangle$ for energies below the soft scale $m_b$. When squaring the amplitude, this gluon operator will eventually become the well-known gluonic parton distribution function of the proton. As a matching coefficient, the form factor for $gg\to h$ may now be computed with on-shell gluons, replacing the gluon operator with gluon polarization vectors. Hence the form factor will feature further divergences, which will later be canceled by the PDFs. For our calculations, we account for this fact by adopting an additional renormalization factor $Z_{gg}^{-1}$. 

We then derived the factorization theorem in terms of renormalized quantities. Since renormalization of the individual component functions and regularization of endpoint divergences within the subtraction scheme does in general not commute due to the occurrence of cutoffs in the integrals of the last term of the factorization formula, we highlight that this is a highly non-trivial achievement. We were able to demonstrate that the additional terms that are introduced by regularizing the renormalized factorization theorem can be absorbed consistently by a redefinition of one of the renormalized hard matching coefficients. The RG evolution equations for the renormalized component functions were presented, as well as the corresponding anomalous dimensions. Furthermore, we solved the RG equations iteratively to predict the leading logarithmic corrections in the $b$-quark induced three-loop amplitude of $gg\to h$ at the order $\mathcal{O}(\alpha_s^3L^k)$, where $k=6,5,4,3$. Eventually, we solved the RG equations for the radiative jet and soft functions to RG-improved leading order. This enabled us to resum the first three leading logarithmic towers (i.e. up to NLL$^\prime$ accuracy) for the $gg\to h$ form factor at all orders of perturbation theory.

We have thus achieved one of the main goals stated in \cite{Liu:2020wbn}, namely the generalization of the SCET analysis of $h\to\gamma\gamma$ to $gg\to h$ as well as a resummation for the three leading logarithmic terms. The resummation of further subleading logarithms that arise in the second and first terms of the factorization formula is left for future work.

\subsubsection*{Acknowledgements}
We would like to thank the Mainz Institute for Theoretical Physics (MITP) for hospitality and support during the workshop {\em Elliptic Integrals in Fundamental Physics\/} (September 12--16, 2022). The work of M.N., M.S.\ and X.W.\ has been supported by the Cluster of Excellence \textit{Precision Physics, Fundamental Interactions, and Structure of Matter} (PRISMA$^+$ EXC 2118/1) funded by the German Research Foundation (DFG) within the German Excellence Strategy (Project ID 39083149). X.W has also been supported by the Deutsche Forschungsgemeinschaft (DFG, German Research Foundation) under Germany's Excellence Strategy – EXC 2094–390783311. Z.L.L.\ is funded by the European Union (ERC, grant agreement No.~101044599, JANUS). Views and opinions expressed are however those of the authors only and do not necessarily reflect those of the European Union or the European Research Council Executive Agency. Neither the European Union nor the granting authority can be held responsible for them.

\begin{appendix}

\renewcommand{\theequation}{A.\arabic{equation}}
\setcounter{equation}{0}

\section{Bare matching coefficients and matrix elements}
\label{app:BareQuant}

In this section, we collect the expressions for the bare quantities of the factorization theorem \eqref{eq:factoGluon}. The hard matching coefficient $H_1^{(0)}$ is given by
\begin{equation}
	\label{eq:H1def}
	H_1^{(0)}=\delta_{a b} T_{F} \frac{y_{b, 0}}{\sqrt{2}} \frac{\alpha_{s, 0}}{\pi}\left[H_{1,0}^{(0)}+\frac{\alpha_{s,0}}{4 \pi} H_{1,1}^{(0)}+\cdots\right] ,
\end{equation}
with
\begin{equation}
	\label{eq:H1cont}
	\begin{aligned}
	H_{1,0}^{(0)}&=(-M_{h}^2-i 0)^{-\epsilon} e^{\epsilon \gamma_{E}}(1-3 \epsilon) \frac{2 \Gamma(1+\epsilon) \Gamma^{2}(-\epsilon)}{\Gamma(3-2 \epsilon)}\,,\\
	H_{1,1}^{(0)}&=(-M_{h}^2-i 0)^{-2\epsilon}\bigg\{C_{F}\bigg[-\frac{1}{2\epsilon^{4}}+\frac{3}{2\epsilon^{3}}-\frac{5 \pi^{2}}{12 \epsilon^{2}}-\frac{1}{\epsilon}\left(\frac{29 \zeta_{3}}{3}+\frac{3 \pi^{2}}{4}+12\right)\\
	&\quad -72- \pi^{2}-19 \zeta_{3}-\frac{3 \pi^{4}}{16}\bigg]+C_{A}\left[-\frac{3}{2\epsilon^{4}}+\frac{1}{\epsilon^{2}}\left(5+\frac{7 \pi ^2}{12}\right)+\frac{18 \zeta_3+14}{\epsilon}\right.\\
 &\quad\left.+20-\frac{2 \pi ^2}{3}+18 \zeta_3+\frac{73 \pi ^4}{240}\right]\bigg\}\,.
	\end{aligned}
\end{equation}
The infinity-bin contribution $\Delta H_1^{(0)}$ in \eqref{eq:factoGluon} and figure~\ref{fig:infinitybin} reads
\begin{equation}
	\label{eq:DeltaH1}
	\begin{aligned}
		\Delta H_1^{(0)}&=-\lim _{\sigma\rightarrow-1} H_{3} \int_{M_{h}}^{\infty} \frac{\df \ell_{-}}{\ell_{-}} \int_{\sigma M_{h}}^{\infty} \frac{\df \ell_{+}}{\ell_{+}} J\left(M_{h} \ell_{-}\right) J\left(-M_{h} \ell_{+}\right) \frac{S_{\infty}\left(\ell_{+} \ell_{-}\right)}{m_{b, 0}}\\
		&=\frac{\alpha_{s,0}T_F\spac\delta_{ab}}{4\pi}\frac{y_{b,0}}{\sqrt{2}}\Bigg\{\frac{(-M_h^2-i0)^{-\epsilon}e^{\epsilon\gamma_E}}{\epsilon^2\Gamma(1-\epsilon)}+\frac{\alpha_{s,0}}{4\pi}(-M_h^2-i0)^{-2\epsilon}e^{2\epsilon\gamma_E}\\
	&\quad\times \Bigg[ C_F\bigg(\frac{3\Gamma(\epsilon)\Gamma(-\epsilon)}{\Gamma(2-2\epsilon)}+\frac{(1+\epsilon)\Gamma^2(-\epsilon)+2\Gamma(-\epsilon)\Gamma(\epsilon)\Gamma(2-2\epsilon)}{2\epsilon^2\Gamma(1-2\epsilon)\Gamma(2-2\epsilon)}\bigg)\\
 &\qquad+C_A\frac{\Gamma(-\epsilon)\Gamma(\epsilon)(3-6\epsilon-2\epsilon^2)}{2\epsilon^2\Gamma(2-2\epsilon)}\Bigg]\Bigg\}\,.
	\end{aligned}
\end{equation}
Similarly, we find
\begin{equation}
	\label{eq:H2series}
	H_{2}^{(0)}(z)= \frac{y_{b, 0}}{\sqrt{2}}\left[H_{2,0}^{(0)}(z)+\frac{\alpha_{s,0}}{4 \pi}H_{2,1}^{(0)}(z)+\cdots\right] ,
\end{equation}
with
\begin{equation}
	\label{eq:H2total}
	\begin{aligned}
		H_{2,0}^{(0)}(z)&=\frac{1}{z}+\frac{1}{1-z}\,,\\
		H_{2,1}^{(0)}(z)&=(-M_{h}^2-i 0)^{-\epsilon}\,e^{\epsilon\gamma_E}\frac{\Gamma(1+\epsilon)\Gamma^2(-\epsilon)}{\Gamma(2-2\epsilon)} \\
		&\quad\times \Bigg\{C_F\left[\frac{2-4\epsilon-\epsilon^2}{z^{1+\epsilon}}-\frac{2(1-\epsilon)^2}{z}
		 -2(1-2\epsilon-\epsilon^2)\frac{1-z^{-\epsilon}}{1-z}\right] \\
		&\quad -C_A\bigg[\frac{2-4\epsilon-\epsilon^2}{z^{1+\epsilon}}-\left(2(1-2\epsilon-\epsilon^2)
		 +\frac{\epsilon^2}{1-\epsilon}\right)\frac{1-z^{-\epsilon}}{1-z}\bigg]+(z\rightarrow 1-z)\Bigg\}\,,
	\end{aligned}
\end{equation}
and
\begin{equation}
	\label{eq:H3}
	H_{3}^{(0)}=-\frac{y_{b,0}}{\sqrt{2}}\left[1-\frac{C_{F} \alpha_{s,0}}{4 \pi}\left(-M_{h}^{2}-i0\right)^{-\epsilon}e^{\epsilon\gamma_E}(1-\epsilon)^{2}\,\frac{2\Gamma(1+\epsilon)\Gamma^2(-\epsilon)}{\Gamma(2-2\epsilon)}\right] .
\end{equation}
for the hard coefficients of the second and third term of the factorization theorem. Note that $H_3$ is the same as in the $h\to\gamma\gamma$ process. The bare soft function of the first term is $S_1^{(0)}=m_{b,0}$ and is exact to all orders of perturbation theory. The soft function of the second term reads
\begin{equation}
	\begin{aligned}
		S_2^{(0)}(z) =m_{b,0}T_F\spac\delta_{ab}\frac{\alpha_{s,0}}{4\pi}\spac\bigg\{
		& 2e^{\epsilon\gamma_E}(m_{b,0}^2)^{-\epsilon}\Gamma(\epsilon) \\
		&+ \frac{\alpha_{s,0}}{4\pi}(m_{b,0}^2)^{-2\epsilon}\bigg[C_FK_{F}(z)+C_AK_{A}(z)+(z\rightarrow 1-z)\bigg]\bigg\}\,,
	\end{aligned}
\end{equation}
with
\begin{equation}
	\label{eq:O2KF}
	\begin{aligned}
		K_F(z)&= \frac{1}{\epsilon^{2}}\left(2L_z+3\right)+\frac{1}{\epsilon}\left(L_z^2-2L_z L_{\bar{z}}-\frac{1}{2}-\frac{\pi^{2}}{3}\right) \\
		&\quad +12 \operatorname{Li}_{3}(z)+2(1-2 z-2 L_z) \operatorname{Li}_{2}(z)+\frac{L_z^3}{3}+2\big[z+L_{\bar{z}}\big] L_z^2 \\
		&\quad +\left(4 \operatorname{Li}_{2}(\bar{z})-L_{\bar{z}}-1-3 z-\frac{\pi^{2}}{3}\right) L_z+3+\frac{\pi^{2}}{3}-8 \zeta_{3}+\mathcal{O}(\epsilon)\,,\\
		K_A(z)&=\frac{-2L_z}{\epsilon^2}+\frac{1}{\epsilon}\left(-L_z^2+\frac{1}{2}\right)-8\operatorname{Li}_{3}(z)+2\operatorname{Li}_{2}(z)\big(z-2L_{\bar{z}}\big)-\frac{L_z^3}{3}\\
		&\quad -4L_z^2L_{\bar{z}}-zL_z^2+\left(1+2z+\frac{\pi^2}{3}\right)L_z+1-\frac{\pi^2}{6}+8\zeta_3+\mathcal{O}(\epsilon)\,.
	\end{aligned}
\end{equation}
The jet function in the third term of the form factor has been derived in \cite{Liu:2021mac} and reads up to NLO
\begin{equation}
	\label{eq:JetNLO}
		J^{(0)}(p^2)=1+\frac{\alpha_{s,0}\left(C_F-C_A\right)}{4\pi}\left(-p^2-i0\right)^{-\epsilon}e^{\epsilon\gamma_E}\frac{\Gamma(1+\epsilon)\Gamma^2(-\epsilon)}{\Gamma(2-2\epsilon)}\left(2-4\epsilon-\epsilon^2\right)\,.
\end{equation}

The soft function of the third term, $S_3^{(0)}$, is more involved than its abelian counterpart due to the additional insertions of two color generators. As shown in section~\ref{subsec:factoGluon}, these lead to the appearance of two semi-finite Wilson lines in the adjoint representation. Therefore, one-loop corrections include exchanges of gluons between Wilson lines in the fundamental and adjoint representation. Feynman diagrams contributing to the soft function are given in figure~\ref{fig:soft}. 
Eventually, the soft function reads
\begin{equation}
	\label{eq:softfinal}
	\begin{aligned}
		S_3^{(0)}(w)=-\frac{T_F\spac\delta_{ab}\,\alpha_{s,0}}{\pi}m_{b,0}\left[S^{(0)}_{a}(w)\,\theta\!\left(w-m_{b,0}^{2}\right)+S^{(0)}_{b}(w) \,\theta\!\left(m_{b,0}^{2}-w\right)\right],
	\end{aligned}
\end{equation} 
with
\begin{align}
	\label{eq:softfinalcont}
	S_a^{(0)}(w)&=\frac{e^{\epsilon \gamma_{E}}}{\Gamma(1-\epsilon)}\left(w-m_{b,0}^{2}\right)^{-\epsilon}\left[1+ \frac{C_{F} \alpha_{s,0}}{4 \pi} 2 e^{\epsilon \gamma_{E}} \frac{3-2 \epsilon}{1-2 \epsilon} \Gamma(1+\epsilon) \frac{\left(m_{b,0}^{2}\right)^{1-\epsilon}}{w-m_{b,0}^{2}}\right]\notag\\
	&\quad +\frac{\alpha_{s,0} C_F}{4\pi}\Bigg\{\left(w-m_{b,0}^{2}\right)^{-2 \epsilon}\left[-\frac{2}{\epsilon^{2}}+\frac{6}{\epsilon}+\frac{2}{\epsilon} \ln \left(1-r\right)+12-\frac{\pi^{2}}{3}\right.\notag \\
	&\quad \left.+\left(24-3 \pi^{2}+\frac{4 \zeta_{3}}{3}\right) \epsilon\right]+\left(m_{b,0}^2\right)^{-2\epsilon}\big[-2 \mathrm{Li}_{2}\left(r\right)+2\left(\ln r+1\right) \ln \left(1-r\right)\notag \\
	&\quad-3 \ln ^{2}\left(1-r\right)\big]\Bigg\}+\frac{\alpha_{s,0} C_A}{4\pi}\Bigg\{\left(w-m_{b,0}^{2}\right)^{-2 \epsilon}\left[\frac{2}{\epsilon^{2}}-\frac{\pi^{2}}{3}-\frac{16}{3} \zeta_{3} \epsilon\right]\notag\\
 &\quad +\left(m_{b,0}^2\right)^{-2\epsilon}\left[4 \mathrm{Li}_{2}\left(r\right)+2 \ln ^{2}\left(1-r\right)\right]\Bigg\}\,,\notag \\
	S_b^{(0)}(w)&=\left(C_{F}-\frac{C_{A}}{2}\right) \frac{\alpha_{s,0}}{4 \pi}\left(m_{b,0}^{2}\right)^{-2 \epsilon}\left[-\frac{4}{\epsilon} \ln \left(1-\frac{1}{r}\right)+6 \ln ^{2}\left(1-\frac{1}{r}\right)\right],
\end{align}
where $r=m_{b,0}^2/w$. 

\section{Renormalization factors}
\renewcommand{\theequation}{B.\arabic{equation}}
\setcounter{equation}{0}
\label{app:RenFac}
Here we collect the renormalization factors of the different component functions. 

The three parameters involved in this process, a) the $b$ quark mass entering the operators, b) the $b$ quark Yukawa coupling entering the hard functions, and c) the QCD coupling constant, are renormalized in the $\overline{\text{MS}}$ subtraction scheme as
\begin{equation}
	\label{eq:paras}
	\begin{aligned} 
		m_{b, 0} =Z_{m} m_{b}(\mu),\quad y_{b, 0} =\mu^{\epsilon} Z_{y} y_{b}(\mu),\quad  \alpha_{s, 0} =\mu^{2 \epsilon} Z_{\alpha_{s}} \alpha_{s}(\mu) ,
	\end{aligned}
\end{equation}
with the renormalization factors
\begin{equation}
	\label{eq:Zparas}
	\begin{aligned}
		Z_{y}=Z_{m}=1-3 C_{F} \frac{\alpha_{s}(\mu)}{4 \pi \epsilon}+\mathcal{O}\left(\alpha_{s}^{2}\right), \quad Z_{\alpha_{s}}=1-\beta_{0} \frac{\alpha_{s}(\mu)}{4 \pi \epsilon}+\mathcal{O}\left(\alpha_{s}^{2}\right).
	\end{aligned}
\end{equation}
Here $\beta_{0}=\frac{11}{3} C_{A}-\frac{4}{3} T_{F} n_{f}$ is the first coefficient of the QCD $\beta$-function, with $n_f= n_b+n_l=5$  being the number of active quark flavors. In order to compare our results in different schemes, we need the following relation between the $b$-quark pole mass and its running mass \cite{Chetyrkin:1997dh,Vermaseren:1997fq}:
\begin{equation}
\label{eq:mpole}
\begin{aligned}
   \frac{m_b(\mu)}{m_b} 
   &= 1+\frac{\alpha_s}{4\pi}C_F(-4+3L_m) \\
   &\quad +\left(\frac{\alpha_s}{4\pi}\right)^2\bigg[C_F^2\bigg(\frac{9 L_m^2}{2}-\frac{21 L_m}{2}+\frac{7}{8}+ (8 \ln 2-5)\pi ^2 - 12 \zeta _3\bigg)\\
		&\quad +C_FC_A\bigg(-\frac{11 L_m^2}{2}+\frac{185 L_m}{6}-\frac{1111}{24}+\frac{4(1-3\ln 2)\pi^2}{3}+6 \zeta _3\bigg)\\
		&\quad +C_FT_F\bigg(2 n_fL_m^2 -\frac{26 n_f}{3} L_m +\frac{\left(143-16 \pi ^2\right) n_b}{6} +\frac{\left(71+8 \pi ^2\right) n_l}{6}\bigg)\bigg]\,,
	\end{aligned}
\end{equation}
with $L_m=\ln (m_b^2/\mu^2)$.

The hard function $H_3(\mu)$ \eqref{eq:H3re} is renormalized by
\begin{equation}
\label{eq:Z33inv}
Z_{33}^{-1}=1+\frac{C_F\,\alpha_s}{4\pi}\bigg[\frac{2}{\epsilon^2}-\frac{2}{\epsilon}\left(L_h-\frac32\right)\bigg]\,.
\end{equation}
 The hard coefficients $\bar{H}_2(\mu)$ and its endpoint counterpart are renormalized in equation \eqref{eq:H2renormalized} and \eqref{eq:H2barrenormalized}. The corresponding renormalization factors are
\begin{align}
		Z_{22}^{-1}(z,z^\prime)&=\delta(z-z^\prime)\nonumber\\
		&+\frac{\alpha_s}{4\pi}\Bigg\{\delta(z-z^\prime)\bigg[(C_F-C_A)\frac{2(L_z+L_{\bar{z}})+3}{\epsilon}+C_A\left(\frac{2}{\epsilon^2}-\frac{2L_h-3}{\epsilon}\right)\bigg]\nonumber\\
		&+\frac{2(C_F-C_A/2)}{\epsilon}z(1-z)\left[\frac{1}{z^{\prime}(1-z)}\frac{\theta\left(z^{\prime}-z\right)}{\left(z^{\prime}-z\right)}+\frac{1}{z(1-z^{\prime})}\frac{\theta\left(z-z^{\prime}\right)}{\left(z-z^{\prime}\right)}\right]_+\Bigg\}\,	\label{eq:Z22},\\
		\braces{Z_{22}^{-1}(z,z^\prime)}&=\delta(z-z^\prime)+\frac{\alpha_s}{4\pi}\Bigg\{\delta(z-z^\prime)\left[(C_F-C_A)\frac{2 L_z+3}{\epsilon}+C_A\left(\frac{2}{\epsilon^2}-\frac{2L_h-3}{\epsilon}\right)\right]\nonumber\\
		&+\frac{\left(2 C_F-C_A\right)}{\epsilon}z\left[\frac{\theta\left(z^{\prime}-z\right)}{z^{\prime}\left(z^{\prime}-z\right)}+\frac{\theta\left(z-z^{\prime}\right)}{z\left(z-z^{\prime}\right)}\right]_+\Bigg\}\,.	\label{eq:Z22invbraces}
\end{align}
At NLO, the renormalization factor for the soft function $S_2(z,\mu)$ \eqref{eq:O2renormalized} is given by
\begin{equation}
	\label{eq:diagZ22}
	\begin{aligned}
		Z_{gg}^{-1}\spac Z_{22}(z,z^\prime) &= \delta(z-z^\prime)+\frac{\alpha_s}{4\pi}\Bigg\{-\frac{3C_F-\beta_0+2(C_F-C_A)\big(L_z+L_{\bar{z}}\big)}{\epsilon}\delta(z-z^\prime)\\
		-&\frac{(2 C_F-C_A)}{\epsilon}z(1-z)\left[\frac{1}{z^{\prime}(1-z)}\frac{\theta\left(z^{\prime}-z\right)}{\left(z^{\prime}-z\right)}+\frac{1}{z(1-z^{\prime})}\frac{\theta\left(z-z^{\prime}\right)}{\left(z-z^{\prime}\right)}\right]_+\Bigg\}\,\\
		Z_{gg}^{-1}\spac Z_{21}(z)&=\frac{T_F\spac\delta_{ab}\alpha_s}{2\pi}\Bigg\{-\frac{1}{\epsilon}+\frac{\alpha_s}{4\pi}\bigg[(C_F-C_A)\bigg(\frac{L_z+L_{\bar{z}}}{\epsilon^2}-\frac{L_z^2+L_{\bar{z}}^2-1}{2\epsilon}\bigg)\\
		&+C_F\frac{2L_z L_{\bar{z}}-6+\pi^2/3}{\epsilon}\bigg]\Bigg\}\,.
	\end{aligned}
\end{equation}
The Jet function and its renormalization have been studied in \cite{Liu:2021mac} in detail. It is renormalized in the convolution sense \eqref{eq:JetRedef}, with the renormalization factor
\begin{equation}
	\label{eq:ZJNLO}
	\begin{aligned}
		Z_{J}\!\left(y p^{2}, x p^{2}\right)&=\left[1+\frac{(C_{F}-C_A) \alpha_{s}}{2\pi}\left(-\frac{1}{\epsilon^{2}}+\frac{L_p}{\epsilon} \right)\right] \delta(y-x)+\frac{(2 C_F-C_A) \alpha_{s}}{4\pi\epsilon}\,\Gamma(y,x) \,.
	\end{aligned}
\end{equation}   
Here $L_p=\ln(-p^2/\mu^2)$, and $\Gamma(y,x)$ is the Lange-Neubert kernel introduced in \eqref{eq:LN}. The plus-distribution is defined such that when $\Gamma(x,y)$ is to be integrated with a function $f(x)$, one has to replace $f(x)\to f(x)-f(y)$ under the integral. Note that the local and the non-local term do not share the same color factor.

Since the bare soft function $S_1^{(0)}\equiv m_{b,0}$, it is renormalized multiplicatively by the renormalization factor of the quark mass. This requires that
	\begin{equation}
		\label{eq:O1re}
		S_1(\mu) = Z_{gg}^{-1}\spac Z_{11} S_1^{(0)} \,, \quad\text{with}\quad 
		Z_{11} = Z_{gg}\spac Z_{m}^{-1} \,.
	\end{equation}

\section{\boldmath Anomalous dimensions and RG functions}
\renewcommand{\theequation}{C.\arabic{equation}}
\setcounter{equation}{0}
\label{app:ADRG}
\stoptocwriting
\subsection{Cusp anomalous dimension}
\label{subapp:cusp}
The cusp anomalous dimension in the fundamental and adjoint representation up to two-loop order is expanded in perturbation theory as
\begin{equation}
    \label{eq:cuspexp}
	\Gamma_\text{cusp}^R(\alpha_s)=\Gamma^R_0\frac{\alpha_s}{4\pi}+\Gamma^R_1\left(\frac{\alpha_s}{4\pi}\right)^2+\dots\,,
\end{equation}
where the superscript $R$ refers to the $SU(N)$ representation. In the case
of QCD, the relevant representations are fundamental ($R = F$) and adjoint
($R = A$). In the $\overline{\text{MS}}$ renormalization scheme the
expansion coefficients in the respective representation are given by
\cite{Korchemsky:1987wg}
\begin{equation}
	\label{eq:cusp}
	\begin{aligned}
	\Gamma_{\text {cusp }}^{R}(\alpha_s)=4C_{R} \left\{\frac{\alpha_{s}}{4\pi}+\left(\frac{\alpha_{s}}{4\pi}\right)^{2}\left[C_{A}\left(\frac{67}{9}-\frac{\pi^{2}}{3}\right)-\frac{20}{9} n_{f} T_{F}\right]+\cdots\right\},
		\end{aligned}
\end{equation}
where $C_R=C_F$ for the fundamental representation while $C_R=C_A$ for the adjoint representation. We introduce the short-hand notations $\Delta\Gamma_0$ and $\Delta\Gamma_1$ which represent the difference of the cusp anomalous dimensions at leading and next-to-leading order
\begin{equation}
	\label{eq:DeltaGamma}
	\begin{aligned}
	\Delta\Gamma=&\Delta\Gamma_0\frac{\alpha_s}{4\pi}+\Delta\Gamma_1\left(\frac{\alpha_s}{4\pi}\right)^2\\
	=&4\left(C_F-C_A\right)\left\{\frac{\alpha_{s}}{4\pi}+\left(\frac{\alpha_{s}}{4\pi}\right)^{2}\left[C_{A}\left(\frac{67}{9}-\frac{\pi^{2}}{3}\right)-\frac{20}{9} n_{f} T_{F}\right]+\cdots\right\}.
	\end{aligned}
\end{equation}

\subsection[Anomalous dimension \texorpdfstring{$\gamma_{gg}$}{Lg}]{Anomalous dimension \texorpdfstring{\boldmath$\gamma_{gg}$}{Lg}}
\label{subapp:gammagg}
The anomalous dimension $\gamma_{gg}$ is associated with the renormalization factor $Z_{gg}$ of the two-gluon operator $O_{gg}$.\footnote{The two-gluon operator is renormalized by $\langle O_{gg}(\mu)\rangle=Z_{gg}\langle O_{gg}^{(0)}\rangle$, hence the renormalized form factor reads $F_{gg}(\mu)=Z_{gg}^{-1} F_{gg}^{(0)}$.} 
To all orders of perturbation theory, it is given by \cite{Becher:2009qa}
\begin{equation}
\label{eq:AGgg}
   \gamma_{gg} = \Gamma_{\text{cusp}}^A(\alpha_s)\spac L_h + 2\gamma_g
   = \frac{\alpha_s}{4\pi}\spac \big( 4C_AL_h-2\beta_0 \big) + \mathcal{O}(\alpha_s^2)\,.
\end{equation}
Here, $\gamma_g$ is the anomalous dimension associated with the gluon wave function renormalization. At two-loop order, it reads \cite{Becher:2009qa}
\begin{equation}
	\label{eq:gammag}
	\begin{aligned}
		\gamma_g &= \frac{\alpha_s}{4\pi}(-\beta_0)+\left(\frac{\alpha_s}{4\pi}\right)^2\left[\left(-\frac{692}{27}+\frac{11 \pi^{2}}{18}+2 \zeta_{3}\right) C_{A}^{2}\right.\\
  &\quad \left.+\left(\left(\frac{256}{27}-\frac{2 \pi^{2}}{9}\right) C_{A}+4 C_{F}\right) T_{F} n_{f}\right].
	\end{aligned}
\end{equation}

\subsection{Anomalous dimensions of component functions}
The renormalization factor of the soft function $S_1(\mu)$ is the same as for the quark mass, and so is its anomalous dimension 
\begin{equation}
    \label{eq:massAD}
	\gamma_{11} - \gamma_{gg} = -\gamma_m
	= \frac{3C_F\alpha_s}{2\pi} + \mathcal{O}(\alpha_s^2) \,.
\end{equation}
The diagonal and off-diagonal elements involved in the RG equation of $S_2(z,\mu)$ and its endpoint region counterpart are
\begin{equation}
  \label{eq:ADdiag}
  \begin{aligned}
  	\gamma_{22}(z,z^\prime) - \gamma_{gg}\spac\delta(z-z') 
	&=-\frac{\alpha_s}{4\pi}\Bigg\{\Big[4(C_F-C_A)\big(L_z+L_{\bar{z}}\big)+6C_F-2\beta_0\Big]\delta(z-z^\prime)\\
  	&\quad +4\left(C_F-\frac{C_A}{2}\right)z\bar{z}\left[\frac{1}{z^{\prime}\bar{z}}\frac{\theta\left(z^{\prime}-z\right)}{\left(z^{\prime}-z\right)}+\frac{1}{z\bar{z}'}\frac{\theta\left(z-z^{\prime}\right)}{\left(z-z^{\prime}\right)}\right]_+\Bigg\} \,, \\
  	\braces{\gamma_{22}(z,z^\prime)} - \gamma_{gg}\spac\delta(z-z') 
   &=-\frac{\alpha_s}{4\pi}\Bigg\{\Big[4(C_F-C_A)L_z+6C_F-2\beta_0\Big]\delta(z-z^\prime),\\
  	&\quad +4\left(C_F-\frac{C_A}{2}\right)z\left[\frac{\theta\left(z^{\prime}-z\right)}{z^{\prime}\left(z^{\prime}-z\right)}+\frac{\theta\left(z-z^{\prime}\right)}{z\left(z-z^{\prime}\right)}\right]_+\Bigg\} \,,
  \end{aligned}
\end{equation}
and
\begin{equation}
	\label{eq:ADoffdiag}
	\begin{aligned}
		\gamma_{21}(z)&=\frac{T_F\spac\delta_{ab} \alpha_s}{\pi}\Bigg\{-1+\frac{\alpha_s}{4\pi}\bigg[(C_F-C_A)\left(1-L_z^2-L_{\bar{z}}^2\right)\\
		&\hspace{2.5cm}\mbox{}+C_F\left(4L_zL_{\bar{z}}-12+\frac{2\pi^2}{3}\right)\bigg]\Bigg\} \,,\\
		\braces{\gamma_{21}(z)}&=\frac{T_F\spac\delta_{ab} \alpha_s}{\pi}\Bigg\{-1+\frac{\alpha_s}{4\pi}\bigg[(C_F-C_A)\left(1-L_z^2\right)+C_F\left(\frac{2\pi^2}{3}-12\right)\bigg]\Bigg\} \,.
	\end{aligned}
\end{equation}
The anomalous dimension for $H_3(\mu)$ is given by
\begin{equation}
	\label{eq:gamma33}
	\begin{aligned}
		\gamma_{33}=\Gamma_{\text{cusp}}^F(\alpha_s)L_h+\gamma_H(\alpha_s)=\frac{C_{F} \alpha_{s}}{\pi}\left(L_{h}-\frac{3}{2}\right)+\mathcal{O}(\alpha_s^2)\,,
	\end{aligned}
\end{equation}
where $\gamma_H=2\gamma_q$, and its expression is known up to three loops \cite{Moch:2005id,Becher:2006mr,Becher:2009qa}. The anomalous dimensions for the jet and soft function $S_3$ in the third term read
\begin{equation}
	\label{eq:ADJetandSoft}
	\begin{aligned}
	\gamma_{J}\!\left(p^{2}, x p^{2}\right)&=\frac{\alpha_s}{\pi}\left[(C_F-C_A)L_p\delta(1-x)+\left(C_F-\frac{C_A}{2}\right)\Gamma(1,x)\right] + \mathcal{O}(\alpha_s^2) \,, \\
  	\gamma_S(w,w^\prime)&=-\frac{\alpha_s}{\pi}\Bigg\{\left[(C_F-C_A)L_w+\frac{3C_F-\beta_0}{2}\right]\delta(w-w^\prime)\\
   &\quad +2\left(C_F-\frac{C_A}{2}\right)w\Gamma(w,w^\prime)\Bigg\} + \mathcal{O}(\alpha_s^2) \,.
	\end{aligned}
\end{equation}
These results satisfy the non-trivial relation \eqref{eq:ADrelation}.

\subsection{RG functions}
\label{subapp:RGfuncs}

The RG functions used in section \ref{sec:resum} are defined as 
\begin{equation}
    \label{eq:RGfunctions}
	\begin{aligned}
	S_{V}(\nu,\mu) &= -\int_{\alpha_s(\nu)}^{\alpha_s(\mu)}\df\alpha\,\frac{\gamma_V(\alpha)}{\beta(\alpha)}\int_{\alpha_s(\nu)}^{\alpha}\frac{\df\alpha^\prime}{\beta(\alpha^\prime)} \,,\\
	 a_{ V}(\nu,\mu) &=-\int_{\alpha_s(\nu)}^{\alpha_s(\mu)}\,\df\alpha\frac{ \gamma_V(\alpha)}{\beta(\alpha)}\,,
	\end{aligned}
\end{equation}
with $\gamma_V$ the respective anomalous dimension. In order to derive the RG-improved solution of the soft function at the leading order, we need to solve the integrals up to the leading order. We find
\begin{equation}
	\label{eq:RGfunctionsLO}
	\begin{aligned}
		a^{(0)}_{V}(\nu,\mu)&=\frac{\gamma_{V,0}}{2\beta_0}\ln r\,,\\
		S^{(0)}_{V}(\nu,\mu)&=\frac{\gamma_{V,0}}{4\beta_0^2}\left[\frac{4 \pi}{\alpha_{s}\left(\nu\right)}\left(1-\frac{1}{r}-\ln r\right)+\left(\frac{\gamma_{V,1}}{\gamma_{V,0}}-\frac{\beta_{1}}{\beta_{0}}\right)(1-r+\ln r)+\frac{\beta_{1}}{2 \beta_{0}} \ln ^{2} r\right],
	\end{aligned}
\end{equation}
where $r=\alpha_s(\mu)/\alpha_s(\nu)$.
\resumetocwriting

\section{Higher-order logarithmic terms in the component functions}
\renewcommand{\theequation}{D.\arabic{equation}}
\setcounter{equation}{0}
\label{app:higherlogterms}
\stoptocwriting
In this section we collect our predictions for the leading logarithmic corrections in higher loop order of the component functions. This is achieved by iteratively solving the RG equations in section~\ref{sec:RG}. Eventually, inserting these expressions into the factorization formula, we are able to predict the leading logarithmic terms in the three-loop expression of the form factor in section~\ref{sec:LL}.
\subsection{Higher-order logarithms in the jet and soft functions}
The jet function has been calculated exactly at the two-loop level in \cite{Liu:2021mac} and reads
\begin{equation}
	\label{eq:JetaNNLO}
	\begin{aligned}
		J(p^2,\mu)=1+\frac{\alpha_s}{4\pi}\Big[\cdots\Big]+\left(\frac{\alpha_s}{4\pi}\right)^2\Big[C_F^2\bar{K}_{FF}+C_FC_A\bar{K}_{FA}+C_A^2\bar{K}_{AA}
		\\
		+C_FT_Fn_f\bar{K}_{Fn_f}+C_AT_Fn_f\bar{K}_{An_f}\Big]\,,
	\end{aligned}
\end{equation} 
with
\begin{equation}
	\label{eq:Kbar}
	\begin{aligned}
		\bar{K}_{FF}&=\frac{L_p^4}{2}-\left(1+\frac{\pi ^2}{6}\right) L_p^2+\left(4 \zeta _3+\pi ^2\right) L_p+\frac{3}{2}-\frac{\pi ^2}{3}-39 \zeta _3+\frac{119 \pi ^4}{360}\,,\\
		\bar{K}_{FA}&=-L_p^4-\frac{11 L_p^3}{9}+\frac{85 L_p^2}{9}-\left(\frac{305}{27}+\frac{\pi ^2}{2}+4 \zeta _3\right) L_p-\frac{317}{162}-\frac{65 \pi ^2}{54}+\frac{793 \zeta _3}{18}-\frac{143 \pi ^4}{360}\,,\\
		\bar{K}_{AA}&=\frac{L_p^4}{2}+\frac{11 L_p^3}{9}-\left(\frac{76}{9}-\frac{\pi ^2}{6}\right) L_p^2+\left(\frac{296}{27}-\frac{11 \pi ^2}{18}\right) L_p+\frac{154}{81}+\frac{85 \pi ^2}{54}-\frac{49 \zeta _3}{18}+\frac{\pi ^4}{15}\,,\\
		\bar{K}_{Fn_f}&=\frac{4 L_p^3}{9}-\frac{20 L_p^2}{9}+\frac{76 L_p}{27}+\frac{14}{81}+\frac{5 \pi ^2}{27}+\frac{8 \zeta _3}{9}\,,\\
		\bar{K}_{An_f}&=-\frac{4L_p^3}{9}  +\frac{20 L_p^2}{9}+\left(\frac{2 \pi ^2}{9}-\frac{58}{27}\right) L_p-\frac{275}{81}-\frac{10 \pi ^2}{27}-\frac{50 \zeta _3}{9}\,.
	\end{aligned}
\end{equation}

The computation of the leading logarithmic behavior of the soft function $S_2(z,\mu)$ and the endpoint-region counterpart $\braces{S_2(z,\mu)}$ requires knowledge of the leading order anomalous dimension. To calculate also sub-leading logarithmic terms would necessitate the anomalous dimension at higher loop order, which is currently unknown. We obtain
\begin{equation}
	\label{eq:O2aNNLO}
	\begin{aligned}
		S_2(z,\mu)  &=\frac{T_F\spac\delta_{ab}\alpha_s}{2\pi}m_b(\mu)g_\perp^{\mu\nu}\Bigg\{-L_m+\frac{\alpha_s}{4\pi}\Big[\cdots\Big]+\left(\frac{\alpha_s}{4\pi}\right)^2\Big[c_3(z)L_m^3+\mathcal{O}(L_m^2)\Big]\Bigg\}\,,\\
		\braces{S_2(z,\mu) } &=\frac{T_F\spac\delta_{ab}\alpha_s}{2\pi}m_b(\mu)g_\perp^{\mu\nu}\Bigg\{-L_m+\frac{\alpha_s}{4\pi}\Big[\cdots\Big]+\left(\frac{\alpha_s}{4\pi}\right)^2\Big[d_3(z)L_m^3+\mathcal{O}(L_m^2)\Big]\Bigg\}\,,
	\end{aligned}
\end{equation}
with
\begin{equation}
	\label{eq:O2aNNLOcont}
	\begin{aligned}
		c_3(z)&=-C_F^2\left[\frac{2L_z^2}{3}+4L_z+3 \right]+C_FC_A\left[\frac{4L_z^2}{3}+\frac{L_zL_{\bar{z}}}{3}+4L_z\right]\\
		&-C_A^2\left[\frac{2L_z^2}{3}+\frac{L_zL_{\bar{z}}}{3}\right]-\beta_0\Big[\frac{C_F-C_A}{3}L_z+\frac{C_F}{2}\Big]+(z\leftrightarrow 1-z )\,,\\
		d_3(z)&=-C_F^2\frac{2\big(L_z+3\big)^2}{3}+C_FC_A\left[\frac{4L_z^2}{3}+4L_z\right] -C_A^2\frac{2L_z^2}{3}-\beta_0\Big[\frac{C_F-C_A}{3}L_z+C_F\Big]\,.
	\end{aligned}
\end{equation}

The soft function $S_3$ is parametrized as
\begin{equation}
	\label{eq:aNNLOSoft}
	\begin{aligned}
	    S_{a}(w, \mu)&=1+\frac{\alpha_{s}}{4\pi}\Big[\cdots\Big]+\left(\frac{\alpha_{s}}{4\pi}\right)^2\Big[r_4L_w^4+r_3L_w^3+r_2L_w^2+r_1L_w+\mathcal{O}(L_w^0)\\
	    &\hspace{3cm}\mbox{}+s_{3 a}(\hat{w}) L_{m}^{3}+s_{2 a}(\hat{w}) L_{m}^{2}+s_{1 a}(\hat{w}) L_{m}+\mathcal{O}\left(L_{m}^{0}\right)\Big]\,,\\
		S_{b}(w, \mu)&=\frac{\alpha_{s}}{4\pi}\Big[\cdots\Big]+\left(\frac{\alpha_{s}}{4 \pi}\right)^{2}\Big[s_{3 b}(\hat{w}) L_{m}^{3}+s_{2 b}(\hat{w}) L_{m}^{2}+s_{1 b}(\hat{w}) L_{m}+\mathcal{O}\left(L_{m}^{0}\right)\Big]\,.
	\end{aligned}
\end{equation}
and the coefficient functions read
\begin{equation}
	\label{eq:aNNLOSoftari}
	\begin{aligned}
		r_4&=\frac{(C_F-C_A)^2}{2}\,,\\
		r_3&=(C_F-C_A)\left(6C_F+\frac{\beta_0}{3}\right)\,,\\
		r_2&=C_F^2\left(6+\frac{ \pi ^2}{2}\right) +C_FC_A\frac{140}{9}+ C_A^2\left(\frac{67}{9}-\frac{\pi ^2}{2}\right)-\frac{16C_F+20C_A}{9}T_Fn_f\,,\\
		r_1&=-C_F^2(75-3\pi^2)-C_FC_A\left(\frac{1297}{27}-\frac{29 \pi ^2}{9}+14 \zeta_3\right)-C_A^2\left(\frac{404}{27}-14 \zeta_3\right)\\
		&+C_FT_Fn_f\left(\frac{428}{27}-\frac{4 \pi ^2}{9}\right)+C_AT_Fn_f\frac{112}{27}\,,
	\end{aligned}
\end{equation}
\begin{equation}
	\label{eq:aNNLOSoftasia}
	\begin{aligned}
		s_{3a}&=4\left(C_F-C_A\right)\left(C_F-\frac{C_A}{2}\right)\ln\left(1-\hat{\omega}^{-1}\right)\,,\\
		s_{2a}&=2C_F^2\left[\ln \left(1-\hat{\omega}^{-1}\right)\left(14 +10 \ln \left(1-\hat{\omega}^{-1}\right)+9 \ln \hat{w} \right)+5 \operatorname{Li}_{2}\left(\hat{\omega}^{-1}\right)\right]\\
		-&2C_FC_A\left[\ln \left(1-\hat{\omega}^{-1}\right)\left(8 +11 \ln \left(1-\hat{\omega}^{-1}\right)+11 \ln \hat{w} \right)+7 \operatorname{Li}_{2}\left(\hat{\omega}^{-1}\right)\right]\\
		+&6C_A^2\left[\ln \left(1-\hat{\omega}^{-1}\right)\left( \ln \left(1-\hat{\omega}^{-1}\right)+ \ln \hat{w} \right)+ \operatorname{Li}_{2}\left(\hat{\omega}^{-1}\right)\right]\\
		+&2\beta_0\left(C_F-\frac{C_A}{2}\right)\ln\left(1-\hat{\omega}^{-1}\right),
	\end{aligned}
\end{equation}
\begin{equation}
	\label{eq:aNNLOSoftasib}
	\begin{aligned}
		s_{3b}&=-4(C_F-C_A)\left(C_F-\frac{C_A}{2}\right)\ln(1-\hat{w})\,,\\
		s_{2b}&=-\left(C_F-\frac{C_A}{2}\right)\Big[C_F\Big(\ln(1-\hat{w})\big(24-4\ln\hat{\omega}+20\ln(1-\hat{w})\big)\\
		&+4\operatorname{Li}_2(\hat{\omega})\Big)-12C_A\ln^2(1-\hat{w})+2\beta_0\ln(1-\hat{w})\Big]\,.
	\end{aligned}
\end{equation}
Note that since $s_{3b}(\hat{w})$, $s_{2b}(\hat{w})$, $s_{1b}(\hat{w})\rightarrow 0$ when $\hat{w}\rightarrow 0$, at order $\mathcal{O}(\alpha_s^3)$ the leading logarithms in the full form factor will not feature contributions from $S_b(w,\mu)$.

In order to predict the full logarithmic behavior of $S_3$ at three loops, the two-loop anomalous dimension $\gamma_S$ would be needed. Using equation~\eqref{eq:ADrelation} it can be inferred from the jet function anomalous dimension. Thus we write
\begin{equation}
\label{eq:ADsoftexpanded}
\begin{aligned}
    \gamma_S(w,w^\prime)&=-\left[\left(\Gamma^F_{\text{cusp}}(\alpha_s)-\Gamma^A_{\text{cusp}}(\alpha_s)\right)L_w-\gamma_s(\alpha_s)\right]\delta(w-w^\prime)\\
		&-2\left(\Gamma^F_{\text{cusp}}(\alpha_s)-\frac{\Gamma^A_{\text{cusp}}(\alpha_s)}{2}\right)w\Gamma(w,w^\prime)-2\left(\frac{\alpha_s}{4\pi}\right)^2g\left(\frac{\hat{w}}{w}\right)+\mathcal{O}(\alpha_s^3)\,,
  \end{aligned}
\end{equation}
where $\Gamma^{F/A}_{\text{cusp}}$ is the cusp anomalous dimension up to two-loop order in the fundamental/adjoint representation. Here, $g(x)$ is an unknown non-local kernel function. In the RG equation for the soft function, it will generate a contribution at order $\mathcal{O}(\alpha_s^3)$ when convoluted with the leading order soft function
\begin{equation}
	\label{eq:hx}
	\begin{aligned}
		2\int_0^\infty\df x\, g(x)\theta(\omega/x-m_b^2)=2\int_0^{\hat{\omega}}\df x\,g(x)\equiv G(\hat{\omega}).
	\end{aligned}
\end{equation}
Although the explicit functional form of $g(x)$ is unknown, its integration over the full space, i.e., $G(\infty)$, has been calculated in \cite{Liu:2021mac} by demanding the cancellation of all single $\epsilon$ poles in two loop jet function. It reads
\begin{equation}
	\label{eq:Ginfty}
	\begin{aligned}
		G(\infty)=&C_F^2\big(4\pi^2-16\zeta_3\big)-C_FC_A\left(\frac{62\pi^2}{9}+24\zeta_3\right)-C_A^2\left(\frac{4}{3}-\frac{22\pi^2}{9}-40\zeta_3\right)\\
		&+C_FT_Fn_f\frac{16\pi^2}{9}+C_AT_Fn_f\left(\frac{8}{3}-\frac{8\pi^2}{9}\right)\,.
	\end{aligned}
\end{equation}
Knowing $G(\hat{w})$ only at the limits does not spoil the accuracy of the prediction of the three-loop logarithms in the form factor, since its contributions will only show up at lower logarithmic order. 

\subsection{Higher-order logarithms in the matching coefficients}

The hard function $H_3(\mu)$ is the same as in the photon case, hence its higher-order logarithmic behavior can be found in \cite{Liu:2020wbn}. The hard coefficients $\bar{H}_2(z,\mu)$ and $\braces{\bar{H}_2(z,\mu)}$ can be parameterized as
\begin{equation}
	\label{eq:aNNLOH2}
	\begin{aligned}
		\bar{H}_2(z,\mu)&=\frac{y_b}{\sqrt{2}}\Bigg\{1+\frac{\alpha_s}{4\pi}\Big[\cdots\Big]+\left(\frac{\alpha_s}{4\pi}\right)^2\Big[a_4L_h^4+a_3L_h^3+a_2L_h^2+\mathcal{O}(L_h) \Big] \Big\}\,,\\
		\braces{\bar{H}_2(z,\mu)}&=\frac{y_b}{\sqrt{2}}\Bigg\{1+\frac{\alpha_s}{4\pi}\Big[\cdots\Big]+\left(\frac{\alpha_s}{4\pi}\right)^2\Big[b_4L_h^4+b_3L_h^3+b_2L_h^2+\mathcal{O}(L_h) \Big] \Big\}\,,
	\end{aligned}
\end{equation}
where we find after solving the evolution equations
\begin{equation}
	\label{eq:aNNLOH2cont}
	\begin{aligned}
		a_4&=b_4=\frac{C_A^2}{2}\,,\\
		a_3&=-2C_A(C_F-C_A)\big(L_z+L_{\bar{z}}\big)+\frac{\beta_0C_A}{3}\,,\\
		b_3&=-2C_A(C_F-C_A)L_z+\frac{\beta_0C_A}{3}\,,\\
		a_2&=(C_F-C_A)\Big[2(C_F-C_A)\big(L_z^2+L_{\bar{z}}^2\big)-C_A\big(L_z+L_{\bar{z}}\big)^2\\
		&-\beta_0\big(L_z+L_{\bar{z}}\big)\Big]+C_A\left[\left(\frac{\pi ^2}{6}-\frac{76}{9}\right)C_A+3C_F+\frac{20}{9}T_Fn_f\right] ,\\
		b_2&=(C_F-C_A)\Big[(2C_F-3C_A)\ln^2z-\beta_0\ln z\Big]+C_A\left[\left(\frac{\pi ^2}{6}-\frac{76}{9}\right)C_A+3C_F+\frac{20}{9}T_Fn_f\right] .
	\end{aligned}
\end{equation}
As a consequence of the complex RG equation for $H_1(\mu)$, we can only predict the first two leading logarithms for this hard function. We eventually find
\begin{equation}
	\label{eq:H1aNNLO}
	\begin{aligned}
		H_1(\mu)=\frac{y_b}{\sqrt{2}}\frac{T_F\spac\delta_{ab}\alpha_s}{\pi}\left[-2+\frac{\alpha_s}{4\pi}\Big[\cdots\Big]+\left(\frac{\alpha_s}{4\pi}\right)^2\Big[c_4L_h^4+c_3L_h^3+\mathcal{O}(L_h^2)\Big]\right] ,
	\end{aligned}
\end{equation}
with
\begin{equation}
	\label{eq:H1aNNLOcont}
	\begin{aligned}
		c_4&=-C_A^2\left(1+\frac{\pi^2}{3}\right)+C_FC_A\frac{\pi^2}{3},\\
		c_3&=C_F^2\left(\frac{2\pi ^2}{3}-\frac{16 \zeta_3}{3}\right)-C_FC_A\left(12-\frac{4\pi ^2}{27}+\frac{8 \zeta_3}{3}\right)-C_A^2\bigg(\frac{22}{9}+\frac{22 \pi ^2}{27}\\
		&-12 \zeta_3\bigg)+C_AT_Fn_f\left(\frac{8}{9}+\frac{8\pi^2}{27}\right)-C_FT_Fn_f\frac{8\pi^2}{27}\,.
	\end{aligned}
\end{equation}

\resumetocwriting

\end{appendix}

\newpage
\addcontentsline{toc}{section}{References}
\bibliographystyle{JHEP}
\bibliography{references}

\providecommand{\href}[2]{#2}\begingroup\raggedright\begin{thebibliography}{10}

\bibitem{Czakon:2020vql}
M.~L. Czakon and M.~Niggetiedt, \emph{{Exact quark-mass dependence of the
  Higgs-gluon form factor at three loops in QCD}},
  \href{https://doi.org/10.1007/JHEP05(2020)149}{\emph{JHEP} {\bfseries 05}
  (2020) 149}, [\href{https://arxiv.org/abs/2001.03008}{{\ttfamily
  2001.03008}}].

\bibitem{ParticleDataGroup:2022pth}
{\scshape Particle Data Group} collaboration, R.~L. Workman et~al.,
  \emph{{Review of Particle Physics}},
  \href{https://doi.org/10.1093/ptep/ptac097}{\emph{PTEP} {\bfseries 2022}
  (2022) 083C01}.

\bibitem{Aparisi:2021tym}
J.~Aparisi et~al., \emph{{mb at mH: The Running Bottom Quark Mass and the Higgs
  Boson}}, \href{https://doi.org/10.1103/PhysRevLett.128.122001}{\emph{Phys.
  Rev. Lett.} {\bfseries 128} (2022) 122001},
  [\href{https://arxiv.org/abs/2110.10202}{{\ttfamily 2110.10202}}].

\bibitem{Liu:2019oav}
Z.~L. Liu and M.~Neubert, \emph{{Factorization at subleading power and
  endpoint-divergent convolutions in $h\to\gamma\gamma$ decay}},
  \href{https://doi.org/10.1007/JHEP04(2020)033}{\emph{JHEP} {\bfseries 04}
  (2020) 033}, [\href{https://arxiv.org/abs/1912.08818}{{\ttfamily
  1912.08818}}].

\bibitem{Liu:2020wbn}
Z.~L. Liu, B.~Mecaj, M.~Neubert and X.~Wang, \emph{{Factorization at subleading
  power and endpoint divergences in $h\to\gamma\gamma$ decay. Part II.
  Renormalization and scale evolution}},
  \href{https://doi.org/10.1007/JHEP01(2021)077}{\emph{JHEP} {\bfseries 01}
  (2021) 077}, [\href{https://arxiv.org/abs/2009.06779}{{\ttfamily
  2009.06779}}].

\bibitem{Liu:2020tzd}
Z.~L. Liu, B.~Mecaj, M.~Neubert and X.~Wang, \emph{{Factorization at subleading
  power, Sudakov resummation, and endpoint divergences in soft-collinear
  effective theory}},
  \href{https://doi.org/10.1103/PhysRevD.104.014004}{\emph{Phys. Rev. D}
  {\bfseries 104} (2021) 014004},
  [\href{https://arxiv.org/abs/2009.04456}{{\ttfamily 2009.04456}}].

\bibitem{Bauer:2000yr}
C.~W. Bauer, S.~Fleming, D.~Pirjol and I.~W. Stewart, \emph{{An Effective field
  theory for collinear and soft gluons: Heavy to light decays}},
  \href{https://doi.org/10.1103/PhysRevD.63.114020}{\emph{Phys. Rev. D}
  {\bfseries 63} (2001) 114020},
  [\href{https://arxiv.org/abs/hep-ph/0011336}{{\ttfamily hep-ph/0011336}}].

\bibitem{Bauer:2001yt}
C.~W. Bauer, D.~Pirjol and I.~W. Stewart, \emph{{Soft collinear factorization
  in effective field theory}},
  \href{https://doi.org/10.1103/PhysRevD.65.054022}{\emph{Phys. Rev. D}
  {\bfseries 65} (2002) 054022},
  [\href{https://arxiv.org/abs/hep-ph/0109045}{{\ttfamily hep-ph/0109045}}].

\bibitem{Bauer:2002nz}
C.~W. Bauer, S.~Fleming, D.~Pirjol, I.~Z. Rothstein and I.~W. Stewart,
  \emph{{Hard scattering factorization from effective field theory}},
  \href{https://doi.org/10.1103/PhysRevD.66.014017}{\emph{Phys. Rev. D}
  {\bfseries 66} (2002) 014017},
  [\href{https://arxiv.org/abs/hep-ph/0202088}{{\ttfamily hep-ph/0202088}}].

\bibitem{Beneke:2002ph}
M.~Beneke, A.~P. Chapovsky, M.~Diehl and T.~Feldmann, \emph{{Soft collinear
  effective theory and heavy to light currents beyond leading power}},
  \href{https://doi.org/10.1016/S0550-3213(02)00687-9}{\emph{Nucl. Phys. B}
  {\bfseries 643} (2002) 431--476},
  [\href{https://arxiv.org/abs/hep-ph/0206152}{{\ttfamily hep-ph/0206152}}].

\bibitem{Becher:2014oda}
T.~Becher, A.~Broggio and A.~Ferroglia, \emph{{Introduction to Soft-Collinear
  Effective Theory}}, vol.~896.
\newblock Springer, 2015,
  \href{https://doi.org/10.1007/978-3-319-14848-9}{10.1007/978-3-319-14848-9}.

\bibitem{Ebert:2018gsn}
M.~A. Ebert, I.~Moult, I.~W. Stewart, F.~J. Tackmann, G.~Vita and H.~X. Zhu,
  \emph{{Subleading power rapidity divergences and power corrections for
  q$_{T}$}}, \href{https://doi.org/10.1007/JHEP04(2019)123}{\emph{JHEP}
  {\bfseries 04} (2019) 123},
  [\href{https://arxiv.org/abs/1812.08189}{{\ttfamily 1812.08189}}].

\bibitem{Beneke:2019kgv}
M.~Beneke, M.~Garny, R.~Szafron and J.~Wang, \emph{{Violation of the
  Kluberg-Stern-Zuber theorem in SCET}},
  \href{https://doi.org/10.1007/JHEP09(2019)101}{\emph{JHEP} {\bfseries 09}
  (2019) 101}, [\href{https://arxiv.org/abs/1907.05463}{{\ttfamily
  1907.05463}}].

\bibitem{Moult:2019mog}
I.~Moult, I.~W. Stewart and G.~Vita, \emph{{Subleading Power Factorization with
  Radiative Functions}},
  \href{https://doi.org/10.1007/JHEP11(2019)153}{\emph{JHEP} {\bfseries 11}
  (2019) 153}, [\href{https://arxiv.org/abs/1905.07411}{{\ttfamily
  1905.07411}}].

\bibitem{Moult:2019uhz}
I.~Moult, I.~W. Stewart, G.~Vita and H.~X. Zhu, \emph{{The Soft Quark
  Sudakov}}, \href{https://doi.org/10.1007/JHEP05(2020)089}{\emph{JHEP}
  {\bfseries 05} (2020) 089},
  [\href{https://arxiv.org/abs/1910.14038}{{\ttfamily 1910.14038}}].

\bibitem{Beneke:2019oqx}
M.~Beneke, A.~Broggio, S.~Jaskiewicz and L.~Vernazza, \emph{{Threshold
  factorization of the Drell-Yan process at next-to-leading power}},
  \href{https://doi.org/10.1007/JHEP07(2020)078}{\emph{JHEP} {\bfseries 07}
  (2020) 078}, [\href{https://arxiv.org/abs/1912.01585}{{\ttfamily
  1912.01585}}].

\bibitem{Moult:2019vou}
I.~Moult, G.~Vita and K.~Yan, \emph{{Subleading power resummation of rapidity
  logarithms: the energy-energy correlator in $ \mathcal{N} $ = 4 SYM}},
  \href{https://doi.org/10.1007/JHEP07(2020)005}{\emph{JHEP} {\bfseries 07}
  (2020) 005}, [\href{https://arxiv.org/abs/1912.02188}{{\ttfamily
  1912.02188}}].

\bibitem{Wang:2019mym}
J.~Wang, \emph{{Resummation of double logarithms in loop-induced processes with
  effective field theory}},  \href{https://arxiv.org/abs/1912.09920}{{\ttfamily
  1912.09920}}.

\bibitem{Beneke:2020ibj}
M.~Beneke, M.~Garny, S.~Jaskiewicz, R.~Szafron, L.~Vernazza and J.~Wang,
  \emph{{Large-x resummation of off-diagonal deep-inelastic parton scattering
  from d-dimensional refactorization}},
  \href{https://doi.org/10.1007/JHEP10(2020)196}{\emph{JHEP} {\bfseries 10}
  (2020) 196}, [\href{https://arxiv.org/abs/2008.04943}{{\ttfamily
  2008.04943}}].

\bibitem{Beneke:2022obx}
M.~Beneke, M.~Garny, S.~Jaskiewicz, J.~Strohm, R.~Szafron, L.~Vernazza et~al.,
  \emph{{Next-to-leading power endpoint factorization and resummation for
  off-diagonal \textquotedblleft{}gluon\textquotedblright{} thrust}},
  \href{https://doi.org/10.1007/JHEP07(2022)144}{\emph{JHEP} {\bfseries 07}
  (2022) 144}, [\href{https://arxiv.org/abs/2205.04479}{{\ttfamily
  2205.04479}}].

\bibitem{Liu:2020ydl}
Z.~L. Liu and M.~Neubert, \emph{{Two-Loop Radiative Jet Function for Exclusive
  $B$-Meson and Higgs Decays}},
  \href{https://doi.org/10.1007/JHEP06(2020)060}{\emph{JHEP} {\bfseries 06}
  (2020) 060}, [\href{https://arxiv.org/abs/2003.03393}{{\ttfamily
  2003.03393}}].

\bibitem{Liu:2021mac}
Z.~L. Liu, M.~Neubert, M.~Schnubel and X.~Wang, \emph{{Radiative quark jet
  function with an external gluon}},
  \href{https://doi.org/10.1007/JHEP02(2022)075}{\emph{JHEP} {\bfseries 02}
  (2022) 075}, [\href{https://arxiv.org/abs/2112.00018}{{\ttfamily
  2112.00018}}].

\bibitem{Becher:2010tm}
T.~Becher and M.~Neubert, \emph{{Drell-Yan Production at Small $q_T$,
  Transverse Parton Distributions and the Collinear Anomaly}},
  \href{https://doi.org/10.1140/epjc/s10052-011-1665-7}{\emph{Eur. Phys. J. C}
  {\bfseries 71} (2011) 1665},
  [\href{https://arxiv.org/abs/1007.4005}{{\ttfamily 1007.4005}}].

\bibitem{Becher:2011dz}
T.~Becher and G.~Bell, \emph{{Analytic Regularization in Soft-Collinear
  Effective Theory}},
  \href{https://doi.org/10.1016/j.physletb.2012.05.016}{\emph{Phys. Lett. B}
  {\bfseries 713} (2012) 41--46},
  [\href{https://arxiv.org/abs/1112.3907}{{\ttfamily 1112.3907}}].

\bibitem{Chiu:2011qc}
J.-y. Chiu, A.~Jain, D.~Neill and I.~Z. Rothstein, \emph{{The Rapidity
  Renormalization Group}},
  \href{https://doi.org/10.1103/PhysRevLett.108.151601}{\emph{Phys. Rev. Lett.}
  {\bfseries 108} (2012) 151601},
  [\href{https://arxiv.org/abs/1104.0881}{{\ttfamily 1104.0881}}].

\bibitem{Chiu:2012ir}
J.-Y. Chiu, A.~Jain, D.~Neill and I.~Z. Rothstein, \emph{{A Formalism for the
  Systematic Treatment of Rapidity Logarithms in Quantum Field Theory}},
  \href{https://doi.org/10.1007/JHEP05(2012)084}{\emph{JHEP} {\bfseries 05}
  (2012) 084}, [\href{https://arxiv.org/abs/1202.0814}{{\ttfamily 1202.0814}}].

\bibitem{Beneke:2022zkz}
M.~Beneke, M.~Garny, S.~Jaskiewicz, J.~Strohm, R.~Szafron, L.~Vernazza et~al.,
  \emph{{Endpoint factorization and next-to-leading power resummation of gluon
  thrust}}, \href{https://doi.org/10.22323/1.416.0068}{\emph{PoS} {\bfseries
  LL2022} (2022) 068}, [\href{https://arxiv.org/abs/2207.14199}{{\ttfamily
  2207.14199}}].

\bibitem{Liu:2020eqe}
Z.~L. Liu, B.~Mecaj, M.~Neubert, X.~Wang and S.~Fleming, \emph{{Renormalization
  and Scale Evolution of the Soft-Quark Soft Function}},
  \href{https://doi.org/10.1007/JHEP07(2020)104}{\emph{JHEP} {\bfseries 07}
  (2020) 104}, [\href{https://arxiv.org/abs/2005.03013}{{\ttfamily
  2005.03013}}].

\bibitem{Hill:2002vw}
R.~J. Hill and M.~Neubert, \emph{{Spectator interactions in soft collinear
  effective theory}},
  \href{https://doi.org/10.1016/S0550-3213(03)00116-0}{\emph{Nucl. Phys. B}
  {\bfseries 657} (2003) 229--256},
  [\href{https://arxiv.org/abs/hep-ph/0211018}{{\ttfamily hep-ph/0211018}}].

\bibitem{Becher:2009cu}
T.~Becher and M.~Neubert, \emph{{Infrared singularities of scattering
  amplitudes in perturbative QCD}},
  \href{https://doi.org/10.1103/PhysRevLett.102.162001}{\emph{Phys. Rev. Lett.}
  {\bfseries 102} (2009) 162001},
  [\href{https://arxiv.org/abs/0901.0722}{{\ttfamily 0901.0722}}].

\bibitem{Aglietti:2006tp}
U.~Aglietti, R.~Bonciani, G.~Degrassi and A.~Vicini, \emph{{Analytic Results
  for Virtual QCD Corrections to Higgs Production and Decay}},
  \href{https://doi.org/10.1088/1126-6708/2007/01/021}{\emph{JHEP} {\bfseries
  01} (2007) 021}, [\href{https://arxiv.org/abs/hep-ph/0611266}{{\ttfamily
  hep-ph/0611266}}].

\bibitem{Bodwin:2021cpx}
G.~T. Bodwin, J.-H. Ee, J.~Lee and X.-P. Wang, \emph{{Analyticity,
  renormalization, and evolution of the soft-quark function}},
  \href{https://doi.org/10.1103/PhysRevD.104.016010}{\emph{Phys. Rev. D}
  {\bfseries 104} (2021) 016010},
  [\href{https://arxiv.org/abs/2101.04872}{{\ttfamily 2101.04872}}].

\bibitem{Lange:2003ff}
B.~O. Lange and M.~Neubert, \emph{{Renormalization group evolution of the B
  meson light cone distribution amplitude}},
  \href{https://doi.org/10.1103/PhysRevLett.91.102001}{\emph{Phys. Rev. Lett.}
  {\bfseries 91} (2003) 102001},
  [\href{https://arxiv.org/abs/hep-ph/0303082}{{\ttfamily hep-ph/0303082}}].

\bibitem{Rathie}
A.~K. Rathie, \emph{A new generalization of generalized hypergeometric
  functions}, \href{https://doi.org/10.48550/ARXIV.1206.0350}{\emph{Le
  Matematiche} {\bfseries LII} (1997) 297--310},
  [\href{https://arxiv.org/abs/1206.0350}{{\ttfamily 1206.0350}}].

\bibitem{Wang:2021vtp}
X.~Wang, \emph{{Next-to-leading power SCET in Higgs amplitudes induced by light
  quarks}}, \href{https://doi.org/10.21468/SciPostPhysProc.7.040}{\emph{SciPost
  Phys. Proc.} {\bfseries 7} (2022) 040},
  [\href{https://arxiv.org/abs/2110.05174}{{\ttfamily 2110.05174}}].

\bibitem{ResummationPaper}
Z.~L. Liu, B.~Mecaj, M.~Neubert and X.~Wang, \emph{{Resummation of
  $h\to\gamma\gamma$ beyond next-to-leading logarithms}}, {\emph{to be
  published} }.

\bibitem{Chetyrkin:1997dh}
K.~G. Chetyrkin, \emph{{Quark mass anomalous dimension to O (alpha-s**4)}},
  \href{https://doi.org/10.1016/S0370-2693(97)00535-2}{\emph{Phys. Lett. B}
  {\bfseries 404} (1997) 161--165},
  [\href{https://arxiv.org/abs/hep-ph/9703278}{{\ttfamily hep-ph/9703278}}].

\bibitem{Vermaseren:1997fq}
J.~A.~M. Vermaseren, S.~A. Larin and T.~van Ritbergen, \emph{{The four loop
  quark mass anomalous dimension and the invariant quark mass}},
  \href{https://doi.org/10.1016/S0370-2693(97)00660-6}{\emph{Phys. Lett. B}
  {\bfseries 405} (1997) 327--333},
  [\href{https://arxiv.org/abs/hep-ph/9703284}{{\ttfamily hep-ph/9703284}}].

\bibitem{Korchemsky:1987wg}
G.~P. Korchemsky and A.~V. Radyushkin, \emph{{Renormalization of the Wilson
  Loops Beyond the Leading Order}},
  \href{https://doi.org/10.1016/0550-3213(87)90277-X}{\emph{Nucl. Phys. B}
  {\bfseries 283} (1987) 342--364}.

\bibitem{Becher:2009qa}
T.~Becher and M.~Neubert, \emph{{On the Structure of Infrared Singularities of
  Gauge-Theory Amplitudes}},
  \href{https://doi.org/10.1088/1126-6708/2009/06/081}{\emph{JHEP} {\bfseries
  06} (2009) 081}, [\href{https://arxiv.org/abs/0903.1126}{{\ttfamily
  0903.1126}}].

\bibitem{Moch:2005id}
S.~Moch, J.~A.~M. Vermaseren and A.~Vogt, \emph{{The Quark form-factor at
  higher orders}},
  \href{https://doi.org/10.1088/1126-6708/2005/08/049}{\emph{JHEP} {\bfseries
  08} (2005) 049}, [\href{https://arxiv.org/abs/hep-ph/0507039}{{\ttfamily
  hep-ph/0507039}}].

\bibitem{Becher:2006mr}
T.~Becher, M.~Neubert and B.~D. Pecjak, \emph{{Factorization and Momentum-Space
  Resummation in Deep-Inelastic Scattering}},
  \href{https://doi.org/10.1088/1126-6708/2007/01/076}{\emph{JHEP} {\bfseries
  01} (2007) 076}, [\href{https://arxiv.org/abs/hep-ph/0607228}{{\ttfamily
  hep-ph/0607228}}].

\end{thebibliography}\endgroup

\end{document}